\newcommand{\bfR}{{\bf R}}
\newcommand{\bfE}{{\bf E}}
\newcommand{\bfC}{{\bf C}}
\newcommand{\bfN}{{\bf N}}
\newcommand{\bfZ}{{\bf Z}}
\newcommand{\bfB}{{\bf B}}

\newcommand{\ep}{\varepsilon}
\newcommand{\nn}{\nonumber}

\newcommand{\ord}[2]{{\cal O}(#1 ^{#2})}

\newcommand{\LOG}[1]{ \log \left( #1 \right)}

\newcommand{\SCR}[1]{{\mathscr #1}}

\newcommand{\CAL}[1]{{\cal #1}}

\newcommand{\MAT}[1]{\left(\begin{array}{cccccccccc}#1\end{array}\right)}
\newcommand{\J}[1]{\left\langle #1 \right\rangle}
\newcommand{\D}[1]{{\mathscr D}( #1 )}

\newcommand{\COS}[1]{\cos{(#1)}}
\newcommand{\SIN}[1]{\sin{(#1)}}
\newcommand{\TAN}[1]{\tan{(#1)}}

\documentclass[preprint,11pt]{article}

\usepackage[top=30truemm, bottom=30truemm, left=25truemm, right=25truemm]{geometry}

\usepackage{fancyhdr}
\usepackage{amsthm}
\usepackage{amsmath,amssymb,latexsym,amsfonts,mathrsfs}

\theoremstyle{definition}
\newtheorem{Thm}{{\bf Theorem}}[section]

\newtheorem{Lem}[Thm]{{\bf Lemma}}

\newtheorem{Cor}[Thm]{{\bf Corollary}}
\newtheorem{Ass}[Thm]{{\bf Assumption}}

\newtheorem{Rem}[Thm]{{\bf Remark}}

\newcounter{Exami}

\newcommand{\Proof}[2][Proof]{
\begin{proof}[{\bf #1}]
#2
\end{proof}
}

\renewcommand{\Re}{\mathop{\rm Re}}
\renewcommand{\Im}{\mathop{\rm Im}}

\usepackage{amssymb}

\begin{document}



\begin{flushleft}
{ \Large \bf MOURRE THEORY FOR TIME-PERIODIC MAGNETIC FIELDS}
\end{flushleft}
\begin{flushleft}
by {\large Masaki Kawamoto}\\
{Department of Mathematics, Graduate School of Science,
 Tokyo university of science \\  1-3,
Kagurazaka, Shinjyuku-ku, Tokyo, 162-8601, Japan \\ E-mail: mkawa@rs.tus.ac.jp}
\end{flushleft}
\begin{abstract}

We study the Mourre theory for the Floquet Hamiltonian  $\hat{H} = -i \partial
 _t + H(t)$ generated by the time-periodic Hamiltonian $H(t)$ with
 $H(t+T) =H(t)$, which describes the Schr\"{o}dinger
 equations with general time-periodic magnetic fields.
\end{abstract}

\begin{flushleft}
{\em Keywards:}  Mourre Theory, Time-Periodic System, Time-Periodic Magnetic Fields,
 Scattering Theory \\
MSC[81U05]

\end{flushleft}




\section{Introduction}

Mourre theory was firstly considered by E. Mourre \cite{Mo} to prove the absence of
 singular spectrum of Schr\"{o}dinger operators, and this theory played
 very important role in the scattering theory, especially,
 in the many-body quantum scattering theory, see e.g., Derezi\'{n}ski and G\'{e}rard
 [DG]. From 1980's to the present, this theory has been applied to the Schr\"{o}dinger operators with various external
 fields. However, there are no results about the Mourre theory for
 the Schr\"{o}dinger operators with time-periodic magnetic fields as far as we
 know and hence we prove this problem in this paper.

We study the dynamics of a charged particle, which is influenced by time-periodic
 magnetic fields, where we assume that the
charged particle is moving in the plane $\bfR ^2$ in the presence of a time-periodic magnetic field $\bfB
(t) = (0,0,B(t))$ with $B(t+T) = B(t)$, which is always perpendicular to this plane. Then the free Hamiltonian
for this system is given by
\begin{align}\label{2}
H_0(t) = (p-qA(t,x))^2/(2m), \quad 
A(t,x) = (-B(t)x_2 , B(t)x_1)/2, \quad \mbox{on } L^2({\bf R}^2)
\end{align}
where $x=(x_1,x_2) \in \bfR ^2$, 
$p=(p_1, p_2) = -i (\partial_1 , \partial _2)$, $m>0$ and $q \in \bfR \backslash
\{0\}$ are the position, the momentum, the mass and the charge of the
particle, respectively. The intense of the magnetic field at $t$ is denoted by
 $B(t) \in L^{\infty}(\bfR)$. We let the wave function $\psi(t,x)$ is the solution to 
 the following time-dependent Schr\"{o}dinger equations: 
\begin{align}\label{1}
i \partial _t \psi(t,x) = H_0(t) \psi(t,x) , \quad 
\psi(0,x) = \psi_0.
\end{align}
Then by denoting the propagator for $H_0(t)$ by $U_0(t,0)$, we can see the wave function $\psi(t,x)$ of \eqref{1} is
denoted by $\psi(t,x) = U_0(t,0) \psi_0$, where we call a family of unitary operators $\{U_0(t,s) \}_{(t,s) \in \bfR ^2}$ {\em a propagator for} $H_0(t)$,
if the each components $U_0(t,s)$ satisfy 
\begin{align*}
&i \partial _t U_0(t,s) =H_0(t)U_0(t,s) , \quad 
i \partial_s U_0(t,s) = - U_0(t,s) H(s), \\ 
&U_0(t,\theta)U_0(\theta ,s) =U_0(t,s), \quad U_0(s,s)= \mathrm{Id}.
\end{align*} 

Here we introduce the result of a paper Adachi-Kawamoto \cite{AK}, which considers the scattering theory for
 a time-periodic pulsed magnetic field and proves the asymptotic completeness of wave operators, where we let 
 the pulsed magnetic field is the following
\begin{align}\label{pu1}
B(t) = \begin{cases}
B, & t \in [NT, NT +T_B], \\ 
0, & t \in [NT+T_B, (N+1)T],
\end{cases}, \quad 
0<T_B<T, \ N \in \bfZ.
\end{align}
In \cite{AK}, the following lemma was showed and that implies the asymptotic behavior of the charged particle 
with respect to \eqref{pu1} is
classified into the three types accordingly to the $B$, $T_B$ and $T$.
\begin{Lem}
Let  $\tilde{D}= 2\COS{qBT_B/(2m)} - qB(T-T_B)\SIN{qBT_B/(2m)
}/2  $ and  $\tilde{\lambda}_{\pm} := \tilde{D}/2 \pm
 \sqrt{\tilde{D}^2/4 -1}$. Then for all $\phi \in C_0^{\infty} (\bfR
 ^2)$ and $N\gg 1$, 
\begin{align}
\nn &\mbox{if} \  \tilde{D}^2 < 4, \  \left\| 
x U_0 (NT,0) \phi 
\right\|_{(L^2 (\bfR ^2))^2} = \ord{1}{} , \\ 
\nn & \mbox{if} \  \tilde{D}^2 = 4,\     \left\| 
x U_0 (NT,0) \phi 
\right\|_{(L^2 (\bfR ^2))^2} = \ord{N}{1} , \\ 
& \mbox{if} \  \tilde{D}^2 > 4,\  
\label{g1}
  \left\| 
x U_0 (NT,0) \phi 
\right\|_{(L^2 (\bfR ^2))^2} = \ord{|\mu _N|}{}, \quad \mu _N = \tilde{\lambda} _+ ^N -
 \tilde{\lambda} _- ^N ,
\end{align}
hold.
\end{Lem} 
Roughly speaking, the term $\left\| xU_0(t,0)\phi
\right\|_{(L^2(\bfR ^2))^2}$ can be regarded like the position of the
particle at $t$, and hence \eqref{g1} implies that the charged
particle moves out to the origin exponentially in $t$ by the effect of the pulsed magnetic field, where we use that either $|\lambda _+|$ or
$|\lambda _-|$ is larger than $1$ if $\tilde{D}^2 >4$. We call this
phenomena {\em the particle is 
in exponentially scattering state} in this paper.

On the other hand, Korotyaev \cite{Ko} considers the same
Hamiltonian $H_0(t)$ in \eqref{2} for time-periodic magnetic fields with the condition
mentioned later and proves the asymptotic completeness of wave operators
by using the following representation of the free propagator: 
\begin{align*}
U_0(t,0) = e^{-i (y_2'(t)/y_2(t)) x^2} e^{-i y_1(t)y_2(t) p^2} e^{i \Omega (t)
 L}e^{-i \LOG{|y_2 (t)|} A/2} \CAL{S}^{\nu (t)},
\end{align*} 
where, $A=( x \cdot p + p \cdot x )/2$ and $L= x_1p_2-x_2p_1$ are
called a generator of dilation group and the angular momentum of
the charged
particle, respectively. We put $\Omega (t) = \int_0^t qB(s)/(2m) ds $ and define $y_j (t)$, $j \in \{1,2\}$ is the solution to 
\begin{align*}
y_j''(t) + \left(\frac{qB(t)}{2m} \right)^2 =0 , \quad \begin{cases}
y_1 (0) = 0, & y_1'(0) =1, \\ 
y_2 (0) =1, & y_2'(0) = 0,
\end{cases}
\end{align*}
respectively. Moreover $\CAL{S}$ is an unitary operator which satisfies $(\CAL{S} \phi) (x) = (-1)
\phi (-x)$ and $\nu (t)$ is the number of zeros of $\zeta _1(s)$ in $s \in [0,t]$. Korotyaev considers the scattering theory for the case where
$y_j(t)$ is described by  
\begin{align}\label{5}
\begin{cases}
y_1 (t) &= ty_2(t) + \chi _1(t), \\ 
y_2(t) &= \chi_2(t),
\end{cases}  \quad \mbox{ or }
\begin{cases}
y_1 (t) &= e^{\lambda t} \chi_1 (t), \\ 
y_2(t)  &= e^{-\lambda t}\chi_2(t),
\end{cases}
\end{align} 
for $\lambda \in \bfR$ and periodic or antiperiodic functions
$\chi_1$ and $\chi_2$. The case where $y_j (t)$ is written by l.h.s. of
\eqref{5} is related to the case where $\tilde{D} = 4$ in the pulsed magnetic
field case, and  $y_j (t)$ is written by r.h.s. of \eqref{5} is related to
\eqref{g1}. 

Now, we investigate what makes the motion of the charged particle \eqref{g1}. Here we firstly introduce the
following Theorem:
\begin{Thm}\label{T1}
Define $\zeta _1(t)$, $\zeta _2 (t)$, $\zeta _1'(t)$ and $\zeta_2'(t)$ as solutions to the following equations; 
\begin{align}\label{4}
\zeta _j ''(t) + \left(
\frac{qB(t)}{2m}
\right)^2 \zeta _j(t) = 0 
, \quad 
\begin{cases}
\zeta _1(0) = 1, & \zeta _1'(0) =0 ,\\
\zeta _2(0) = 0, & \zeta _2'(0) =1,
\end{cases}
\end{align}
respectively, and suppose that these are continuous functions. Let $U_0(t,0)$ be the propagator for $H_0(t)$, and then $U_0(t,0)$ is
 factorized as  
\begin{align}\label{3}
U_0(t,0) = e^{-ia(t)x^2} e^{-ib(t)p^2} e^{i \Omega (t)L}e^{-i c(t) x^2},
\end{align}
where $\Omega (t) = \int_0^tqB(s)/(2m) ds$, $L = x_1p_2 -x_2 p_1$ and 
\begin{align}\label{301}
a(t) = \frac{m}{2}\left(
\frac{1- \zeta _2'(t)}{\zeta _2(t)}
\right), \quad 
b(t) = \frac{\zeta _2(t)}{2m}, \quad 
c(t) =  \frac{m}{2}\left(
\frac{1- \zeta _1(t)}{\zeta _2(t)}
\right).
\end{align}
\end{Thm}
\begin{Rem}
This lemma can be proven without the periodic condition $B(t+T) =B(t)$ (see \S{2}). 
\end{Rem}
M\o ller \cite{Mo} studied the scattering theory for a charged particle influenced by time-periodic electric fields $\bfE(t+T_E) = \bfE(t)$, $\bfE(t) =
(E_1(t), ...., E_n(t))$ with $\int_0^{T_E} \bfE(s) ds \neq 0$, and
noticed that the propagator $U_0^S(t,0)$ for the Hamiltonian which describes this system was expressed by 
\begin{align}\label{1000}
U_0^S(t,s) = \tilde{T}(t) e^{-i(t-s) H_0^S} (\tilde{T} (s))^{\ast},
\end{align}
where $\tilde{T}$ was an unitary operator with $\tilde{T} (t+T_E) = \tilde{T} (t)$, $\tilde{T}(\tilde{T})^{\ast} =
\mathrm{Id}$ and  $H_0^S$ was the time-independent Stark-Hamiltonian. Thus, one
can obtain $U_0^S(NT_E,0) = \tilde{T} (0) e^{-iNT_EH_0^S} (\tilde{T}
(0))^{\ast}$, and see that the asymptotic behavior of the charged particle
is governed only by the propagator $e^{-it H_0^S}$. 
From this
expression, many results such as Mourre theory, propagation estimate,
asymptotic completeness of wave operators and so on are proven by
the almost same way of time-independent case. Thus, the first aim of this
paper is to find the time-independent Hamiltonian such that the
propagator \eqref{3} is decomposed like \eqref{1000}.  

Here, define so-called repulsive Hamiltonian as 
\begin{align}
H_R = p^2 - x^2.
\end{align} 
Then it is known that for all $\psi \in L^2(\bfR ^n)$, a particle $e^{-it(p^2-x^2)} \psi$ is in exponentially scattering state,
see Bony-Carles-H$\ddot{a}$fner-Michel \cite{BCHM}, and hence it can be expected that, in the case where $y_j (t) = e^{- \lambda (-1)^j}
\chi_j (t)$ in \eqref{5} (or $\tilde{D} >4$
in pulsed magnetic case), the free
propagator \eqref{3} is decomposed like  
\begin{align*}
U_0(t,s) \sim \tilde{\SCR{J}}(t) e^{-i(t-s) H_R} (\tilde{\SCR{J}} (s) )^{\ast} 
\end{align*} 
for some unitary operator $\tilde{\SCR{J}}$ with $
\tilde{\SCR{J}}(t+T)= \tilde{\SCR{J}} (t) $ and $ \tilde{\SCR{J}}(t)
(\tilde{\SCR{J}} (t) )^{\ast} = \mathrm{Id}$ by the same scheme of
\cite{M}. By the following Lemma \ref{L50}, our expectation will be proven.
\begin{Lem}\label{L50}
Let $D$, $\sigma _D$, $A_D$, $B_D$, $C_D$ and $D_D$ are the followings 
\begin{align}\nn
D &= \zeta _1(T) + \zeta _2'(T),  \quad 
\sigma _D= 
\begin{cases}
0 & D \geq 2 \\
1 & D \leq -2
\end{cases} , \quad 
D_D= \begin{cases}
- \Omega (T)/B_D & D \geq 2 \\ 
-(\Omega (T) + \pi)/B_D & D \leq -2 
\end{cases}, \\ \nn
A_D &= \begin{cases}
a(T) + (a_3 -1)^2/(8a_3b(T)) & D \geq 2 \\ 
a(T) - (b_3+1)^2/(8b_3 b(T)) & D \leq -2 
\end{cases}, \\ 
B_D &= \begin{cases}
2b(T)a_3(\log a_3)/(a_3^2 -1) & D \geq 2 \\ 
-2b(T)b_3(\log b_3)/(b_3^2 -1) & D \leq -2 , 
\end{cases}
 \quad 
C_D =
\begin{cases}
(a_3^2 -1) /(4a_3b(T)) & D \geq 2 \\ 
(b_3^2 -1) /(4b_3b(T)) & D \leq -2
\end{cases} \label{999}
\end{align}
where $\Omega (t) = \int_0^t qB(s)/(2m) ds$, $a(T)$, $b(T)$, $c(T)$ are
 equivalent to those in \eqref{301}. $a_3$ and $b_3$ are given as the solutions of  
\begin{align*}
&a_3^2 - Da_3 +1 =0, \quad a_3 >1, \\ 
&b_3^2 + Db_3 +1=0 , \quad b_3 >1.
\end{align*}
Furthermore, denote 
\begin{align*}
W_0 = (B_D/T) (p^2 - C_D^2 x^2 + D_DL)+ (\pi \sigma _D)/T .
\end{align*}
Then the monodromy operator $U_0(T,0)$ is decomposed into 
\begin{align}\label{6}
U_0(T,0) =  e^{-iA_Dx^2} e^{-iTW_0} e^{iA_Dx^2} , \quad 
U_0(NT,0)= e^{-iA_Dx^2} e^{-iNTW_0} e^{iA_Dx^2}.
\end{align}
$D$ is often called {\em discriminant} of Hill's equation.
\end{Lem} 
Now we consider the asymptotic behavior of a charged particle for the
 case of $D^2 >4$.
 Let us define 
\begin{align*}
x_w(t) \equiv e^{itW_0} xe^{-itW_0}, \quad \tilde{x} _{\omega} (t) \equiv 
e^{-it (B_DD_D/T)L} x_{\omega} (t) e^{it (B_DD_D/T)L}
.
\end{align*} 
Then, the straightforward calculation shows 
\begin{align*}
\tilde{x}''_w(t)  =  \CAL{D}^2 \tilde{x}_w(t), \quad \CAL{D} = 2B_DC_D/T . 
\end{align*}
By using this, $x_w(t)$ can be calculated explicitly, and we have
\begin{align}\label{P1}
{x}_w(t) = (\hat{R} (NB_DD_D)x) \cosh (t \CAL{D}) +  (\hat{R}
 (NB_DD_D)p)(1/C_D)
\sinh (t \CAL{D}),
\end{align}
where we define
\begin{align}\label{302}
\hat{R}(t) =  \MAT{
\COS{t} & \SIN{t} \\ 
- \SIN{t} &\COS{t}
}.
\end{align}
By \eqref{P1}, we notice that the charged particle
is in exponentially scattering state for the case where $D^2 > 4$.

In this paper, we do not deal with the case where 
$D^2 \leq 4$ by the following reasons: \\ 
1. If ${D}^2 <4$ holds, the charged particle never move out the
some compact region by the influence of the magnetic field, and which implies that $L^2(\bfR ^2)
= \SCR{H} _{\mathrm{pp}} (U_0(T,0))$ holds, where $\SCR{H}_{\mathrm{pp}}
(U_0(T,0))$ stands for the pure point spectral subspace associated with $U_0(T,0)$. This result can be proven by
\eqref{P1} with $\cosh (i t \alpha) = \cos (t \alpha)$ and $\sinh (it
\alpha) = i \sin (t \alpha)$. \\
2. Let $D^2=4$. Noting $C_D=0$ if $D^2=4$, denote $\hat{P}_0 = -i \partial _t + p^2 + L$. In order
to consider the Mourre theory for this operator, we
need to obtain the selfadjoint operator $A$ such that $i[\hat{P}_0, A]$ is
a positive valued operator on the support of the energy cut-off. However, even if the case where $A = (x \cdot p)\J{p}^{-2} + \J{p}^{-2} (p
\cdot x)$, which was introduced by Yokoyama \cite{Yo}, to induce the
positivity of commutator $i[\hat{P}_0 ,A]$ is difficult since the
operator $L$ is unbounded operator. In addition to this, the scattering
theory for the case $D^2 =4$ with $n=2$ was not proven in the both of the papers \cite{AK} and \cite{Ko}.

Be based on these reasons, we impose the following
assumption on magnetic field :
\begin{Ass}\label{A1}
Suppose that $\zeta _1(t)$, $\zeta _2 (t)$, $\zeta _1'(t)$ and $\zeta _2'(t)$ in \eqref{4} are continuous functions in $t$ and that 
 the discriminant of the fundamental solutions of Hill's equations \eqref{4}
 satisfy $D^2 = (\zeta _1 (T) + \zeta _2'(T))^2 >4$. Furthermore, $\zeta
 _2(T) \neq 0$ holds.
\end{Ass}
By the virtue of the result of Floquet, see e.g. Magnus and Winkler
\cite{MW}, see also Colombini and Spagnolo
\cite{CS}, the following Lemma holds:
\begin{Lem}
If $D^2 >4$ holds, then the solution $\zeta _0 (t)$ of $\eqref{4}$ is expressed
 by $\zeta _0 (t) = e^{\tilde{\lambda} t} \tilde{\chi} _1 (t) +
 e^{-\tilde{\lambda} t} \tilde{\chi} _2 (t)$ for $\tilde{\lambda} \neq
 0$ and periodic or antiperiodic functions $\tilde{\chi} _1$ and
 $\tilde{\chi} _2$.
\end{Lem}
By virtue of \eqref{6}, $\lambda$ and $\chi _j$ in \eqref{5} will be
 calculated more precisely in \S{3}. What we emphasize here is that,
 we study the 
 Mourre theory by using four quantities $\zeta _j (T)$ and $\zeta _j'(T)$ with $j \in \{1,2\}$. Indeed, as the subconsequent of 
Lemma \ref{L50}, we have the following theorem; 
\begin{Thm}
Suppose Assumption \ref{A1}, then for all $t \in [NT, (N+1)T)$ the solutions of Hill's equation $\zeta _1(t)$ and $\zeta _2 (t)$ can be written as 
\begin{align*}
\nn \zeta _1(t) &= (\cosh (NT \CAL{D}) + (2A_D/C_D) \sinh  (NT \CAL{D}))\zeta _1(t-NT) \\ & \qquad 
+(( C_D \sinh  (NT \CAL{D}) -(4A_D ^2/C_D) \sinh  (NT \CAL{D}))/m)\zeta _2(t-NT), \\
\nn \zeta _2(t) &= ((m/C_D) \sinh  (NT \CAL{D}))\zeta
 _1(t-NT) \\ & \qquad + ((\cosh (NT \CAL{D}) - (2A_D/C_D) \sinh (NT \CAL{D})))\zeta _2(t-NT), \\ 
\nn \zeta'_1(t) &= (\cosh (NT \CAL{D}) + (2A_D/C_D) \sinh  (NT \CAL{D})) \zeta _1'(t-NT) \\ & \qquad + 
 ((C_D \sinh  (NT \CAL{D}) -(4A_D ^2/C_D) \sinh  (NT \CAL{D}))/m) \zeta _2'(t-NT), \\ 
 \zeta'_2(t) &=  ((m/C_D) \sinh  (NT \CAL{D})) \zeta _1'(t-NT) \\ & \qquad + (\cosh (NT \CAL{D}) - (2A_D/C_D) \sinh (NT \CAL{D})) \zeta
 _2'(t-NT). 
\end{align*}
and 
\begin{align*}
\CAL{D} = 2B_D C_D /T
\end{align*}
\end{Thm} 
 Proof of this theorem can be seen in \S{3.3}. Here we remark that the fundamental
discriminant is calculated concretely, see e.g. Hodchstadt \cite{Ho} and Xu
\cite{X}. 
\begin{Rem}
By using $D$, $\zeta _j (T)$ and $\zeta _j '(T)$, the
 conditions of a magnetic field in \cite{Ko} ($=$ r.h.s. of \eqref{5})
 can be written $D >2$, $\zeta _2(T) \neq 0$ and $\zeta _2'(T) = a_3$ or $D < -2$,
 $\zeta _2(T) \neq 0$ and $\zeta _2'(T) = - b _3$, see \S{3}. 
\end{Rem}
Let us define $\SCR{J}_D(t)$ as 
\begin{align}
\SCR{J}_D(t) = U_0(t,0) e^{-i A_Dx^2}e^{it
 {W}_0}.
\end{align}
Then, by \eqref{6},  
\begin{align}\label{8}
\SCR{J}_D(t+T) =U_0(t,0)U_0(T,0) e^{-iA_Dx^2}
e^{i T{W}_0}e^{it{W}_0} = \SCR{J}_D(t)
\end{align}
holds where we use $e^{i 2 \pi L} = \mathrm{Id}$. Here we redefine $U_0 (t,0)$ as 
$$
U_0(t,0) = \SCR{J}_D(t)e^{-i t {W}_0}e^{iA_Dx^2},
$$
and define $\CAL{W}(t,0)$ as a propagator for 
\begin{align}
\tilde{W}(t) = {W}_0 + \SCR{J}_D^{\ast}(t)V \SCR{J}_D(t),
\end{align} 
where $V$ is a multiplication operator by a real-valued function $V(x)$, i.e.,
which satisfies $(V \phi)(x) = V(x) \phi
(x)$ for all $\phi \in L^2(\bfR ^2)$.
Moreover, let us define $U(t,0)$ as 
$$
U(t,0) = \SCR{J}_D(t)\CAL{W}(t,0) e^{iA_Dx^2}.
$$
Then, paying attention to the condition 
\begin{align*}
i \frac{d}{dt} \CAL{W}(t,0) = \tilde{W}(t)\CAL{W}(t,0),
\end{align*}
we have 
\begin{align*}
i \frac{d}{dt} U(t,0) &= 
\left(
\left(i \frac{d}{dt} \SCR{J}_D(t) \right) + \SCR{J}_D(t)({W}_0) \SCR{J}_D(t)^{\ast}
\right) U(t,0) + V U(t,0) \\ 
&= (
 H_0(t) +V) U(t,0).
\end{align*}
Thus, the operator $U(t,0)$ is a propagator for $H(t) = H_0
(t) +V$, and uniqueness and the existence of the propagator $U(t,0)$ are guaranteed
under the following assumption \ref{A4}.
\begin{Ass}\label{A4}
Let $V$ satisfies $(V \phi)(x) = V(x) \phi (x)$ for all $\phi \in
 L^2(\bfR ^2)$, and $V(x) \in C^2(\bfR ^2)$ satisfies that 
for some $\gamma _j >0$, $j \in \{ 0,1,2 \}$, there exists $\tilde{C}_j > 0$ such
 that  
\begin{align*}
| \nabla ^{j} V(x)| \leq \tilde{C}_j \J{x}^{- \gamma _j }, \quad \J{x} = (1+x^2)^{1/2}
\end{align*}
holds.
\end{Ass}
The uniqueness and the existence of $\CAL{W}(t,0)$ are guaranteed under the assumption \ref{A4} too. 
\begin{Rem}
In order to consider the effect of magnetic field, we need to suppose that the charged particle moves in dimension $n=2$ only. But the
 same problems for Harmonic-oscillator with time-periodic coefficients $h(t)$ case,
the Hamiltonian of this system is described by 
\begin{align}
H_{h,0} (t) = p^2/(2m) + h(t)^2x^2/(8m), 
\end{align}
should be considered in general dimension $n$. For $H_{h,0} (t)$, by regarding $L=0$ in $H_0(t)$, above-mentioned
 results Theorem \ref{T1}, Lemma \ref{L50} and Proposition \ref{P1}, and
 the following results Theorem \ref{T2} and Corollary
 \ref{Co1} can be proven by the
 same way under the Assumption \ref{A1} and Assumption \ref{A4} with
 replacement $n=2 \to n=n$ and $qB(t) \to h(t) $.
\end{Rem}
Now we define the Floquet Hamiltonian of time-periodic magnetic
fields. The basic formulation $\hat{P}_0 = -i \partial _t + H_0(t)$ was
considered by \cite{Ko} and \cite{AK}. However, for the sake of constructing the Mourre theory, 
we construct it thorough \eqref{6} with a little technical approach. 

Let ${\bf T}$ be a torus ${\bf T} = {\bf R} / (T {\bf Z})$ and the unitary groups 
$\CAL{L}_0^{\sigma}$ and $\CAL{L}^{\sigma}$, acting on $\SCR{K} := L^2({\bf T};L^2({\bf R}^2))$, as follows:
\begin{align*}
\begin{cases}
(\CAL{L}_0^{\sigma} \phi)(t) &= U_0(t,t-\sigma) \phi(t- \sigma), \\
(\CAL{L}^{\sigma} \phi) (t) &= U(t,t- \sigma) \phi (t-\sigma),
\end{cases}
\quad 
\phi(t) \in \SCR{K}.
\end{align*}
Then $\CAL{L}_0^{\sigma}$ and $\CAL{L}^{\sigma}$ are the unitary operator
on $\SCR{K}$, and satisfy 
\begin{align*}
\begin{cases}
(\CAL{L}_0^{\sigma_1}\CAL{L}_0^{\sigma_2} \phi)(t) &= 
 U_0(t,t- \sigma _2)U_0(t- \sigma _2, t- \sigma _1 - \sigma _2) \phi(t-
 \sigma _1 - \sigma _2) =( \CAL{L}_0^{\sigma _1+ \sigma _2} \phi)(t),\\ 
(\CAL{L}^{\sigma_1}\CAL{L}^{\sigma_2} \phi)(t) &= 
 U(t,t- \sigma _2)U(t- \sigma _2, t- \sigma _1 - \sigma _2) \phi(t-
 \sigma _1 - \sigma _2) =( \CAL{L}^{\sigma _1+ \sigma _2} \phi)(t).
\end{cases}
\end{align*}
Thus, by Stone's theorem, we have selfadjoint operators 
$\hat{H}_0$ and $\hat{H}$ be such that 
\begin{align*}
(e^{-i\sigma \hat{H}_0} \phi)(t) &= (\CAL{L}_0^{\sigma} \phi)(t) = 
U_0(t,t- \sigma)\phi(t- \sigma), \\ 
(e^{-i\sigma \hat{H}} \phi)(t) &= (\CAL{L}^{\sigma} \phi)(t) = 
U(t,t- \sigma)\phi(t- \sigma).
\end{align*} 
Now, by \eqref{8}, ${\SCR{J}_D}(t+T) = {\SCR{J}_D}(t)$ holds, and which implies that, for all $\phi \in \SCR{K}$, $({\SCR{J}_D} \phi) \in 
\SCR{K}$ holds, and hence we have that ${\SCR{J}_D}(t)$ is
an unitary operator on $\SCR{K}$.
Here, we further define $\CAL{M}_0^{\sigma}$ and $\CAL{M}^{\sigma}$ as 
\begin{align*}
\begin{cases}
(\CAL{M}_0^{\sigma} \phi)(t) &= ({\SCR{J}_D}^{\ast} \CAL{L}^{\sigma} _0
{\SCR{J}_D} \phi)(t) = {\SCR{J}_D}^{\ast}(t) e^{-i \sigma \hat{H}_0}({\SCR{J}_D} \phi)(t), \\ 
(\CAL{M}^{\sigma} \phi)(t) &= ({\SCR{J}_D}^{\ast} \CAL{L}^{\sigma} 
 {\SCR{J}_D} \phi)(t) =  {\SCR{J}_D}^{\ast}(t) e^{-i \sigma \hat{H}}({\SCR{J}_D} \phi)(t).
\end{cases} 
\end{align*}
Straightforward calculation shows 
\begin{align*}
\begin{cases}
(\CAL{M}_0^{\sigma} \phi)(t) &= e^{-i \sigma {W}_0} \phi(t- \sigma), \  (=
e^{-it {W}_0}e^{i(t- \sigma){W}_0} \phi(t-\sigma)
) ,\\
(\CAL{M}^{\sigma} \phi)(t) &= \CAL{W}(t,t-\sigma) \phi(t- \sigma).
\end{cases} 
\end{align*}
These equations yield 
two selfadjoint operators $\hat{W}_0$ and $\hat{W}
$ such that 
\begin{align*}
(e^{-i\sigma \hat{W}_0} \phi)(t) &= (\CAL{M}_0^{\sigma} \phi)(t) = 
\CAL{W}_0(t,t- \sigma)\phi(t- \sigma) , \\ 
(e^{-i\sigma \hat{W}} \phi)(t) &= (\CAL{M}^{\sigma} \phi)(t) = 
\CAL{W}(t,t- \sigma)\phi(t- \sigma).
\end{align*} 
By using these equations, we have 
\begin{align*}
e^{-i \sigma \hat{W}_0} = {\SCR{J}_D}^{\ast} e^{-i \sigma \hat{H}_0} {\SCR{J}_D}
 , \quad 
e^{-i \sigma \hat{W}} = {\SCR{J}_D}^{\ast} e^{-i \sigma \hat{H}} {\SCR{J}_D},
\end{align*}
and  
\begin{align*}
{\SCR{J}_D}^{\ast}\hat{H}_0 {\SCR{J}_D} = \hat{W}_0 , \quad 
{\SCR{J}_D}^{\ast}\hat{H}{\SCR{J}_D} = \hat{W}
\end{align*}
Sum of all results, the Floquet Hamiltonian in this model can be described by 
\begin{align*}
\hat{W} = -i \partial _t +W_0 + {\SCR{J}_D}^{\ast} V
 \SCR{J}_D , \quad \mbox{on  }
\SCR{K}
.
\end{align*}
Throughout the followings, we denote $\hat{W}_0 = -i \partial _t +W_0$ and $\hat{V} =  {\SCR{J}_D}^{\ast} V
 \SCR{J}_D
$. The Mourre theory for ${W}_0$ with respect to $L \equiv 0$ was proven by \cite{BCHM}.
The difference between the Hamiltonian we consider and the Hamiltonian
 considered by \cite{BCHM} is whether the potential is a pseudo-differential
 operator or a usual multiplication operator, and hence we need to impose the more stronger
 assumption on $V$ than \cite{BCHM} in order to deduce Mourre estimate.
\begin{Thm}\label{T2}
Suppose Assumption \ref{A1} and Assumption \ref{A4}.  For all 
$\phi \in \SCR{K}$, we define the unitary operators $J$ and 
$K$  as follows
$$(J \phi_{})(t,x) = {\SCR{J}_D}(t) \phi_{}(t,x), \quad (K \phi_{})(t,x) = e^{i(C_D/2)x^2}
 e^{i(1/(4C_D)) p^2} \phi (t,x),$$ and we further define $\varphi \in C_0^{\infty} ({\bf R} \backslash \sigma _{pp}(\hat{H} )
 $, and a conjugate operator $\hat{A}$ as follows
\begin{align*}
\hat{A} = JK \SCR{A}_0 K^{\ast}J^{\ast} , \quad 
\SCR{A}_0 =(T/(4B_DC_D)) (\log \J{x} - \log \J{p}).
\end{align*}
Then, the Mourre estimate holds, i.e., there exist $\delta >0$  and a compact operator $\CAL{K}_{0}$ such that 
\begin{align*}
(i[\hat{H} , \hat{A}] \varphi (\hat{H}) \phi , 
\varphi (\hat{H})  \phi) 
\geq \delta \| \varphi (\hat{H})  \phi \|^2
 + 
( \CAL{K}_{0}  \varphi (\hat{H}) \phi,  \varphi (\hat{H})  \phi)
\end{align*}
holds.
\end{Thm}
\begin{Cor}\label{Co1}
Under assumption \ref{A1} and Assumption \ref{A4}, $\hat{H}$ has at most countable
 pure point spectrum and which singular continuous spectrum is empty.
\end{Cor}
\begin{Rem}
The author could not prove so-called propagation estimate by using the Mourre
estimate and could not include the Columb type potential in Assumption
\ref{A4} by several reasons. 
\end{Rem}

\section{Proof of theorem \ref{T1}}

We find $a(t)$, $b(t)$, $c(t)$ and $d(t)$ be such that  
\begin{align*}
i \frac{d}{dt} e^{-ia(t)x^2} e^{-ib(t)p^2}e^{-id(t) L}e^{-i c(t) x^2}
 = 
H_0(t)e^{-ia(t)x^2} e^{-ib(t)p^2}e^{-id(t) L}e^{-i c(t) x^2} .
\end{align*}
Let $\CAL{K}(t) = e^{-ia(t)x^2} e^{-ib(t)p^2}e^{-id(t) L}e^{-i
c(t) x^2}$, and then 
\begin{align*}
i \frac{d}{dt} \CAL{K}(t) &= \Big\{
a'(t) x^2 +b'(t) (p+2a(t) x)^2 +c'(t)  e^{-ia(t)x^2}(x-2b(t)
 p)^2 e^{ia(t) x^2} + d'(t) L \Big\}\CAL{K}(t) \\ &  
= \Big[2 \big\{4a(t)b'(t) -4b(t) c'(t)(1-4a(t)b(t))
 \big\} A + 
\big\{
b'(t) +4b(t)^2 c'(t)
\big\} p^2 \\ & \qquad + \big\{a'(t) +4a(t)^2 b'(t) + c'(t)(1-4a(t)b(t))
 ^2\big\}x^2 + d'(t)L
 \Big]\CAL{K}(t)
\end{align*}
holds, where $A = x \cdot p + p \cdot x $. This equation yields the following differential equations 
\begin{align*}
\begin{cases}
\mathrm{(eq1)} & b'(t) +4b(t)^2 c'(t)=1/(2m) \\
\mathrm{(eq2)} & 4a(t)b'(t) -4b(t) c'(t) +16 a(t)b(t)^2
 c'(t) =0 \\
\mathrm{(eq3)} & a'(t) +4a(t)^2 b'(t) + c'(t) -8a(t)
 b(t)c'(t) +16 a(t)^2 b(t)^2 c'(t)  =q^2B(t)^2/(8m) \\ 
\mathrm{(eq4)} & d'(t) = - qB(t)/(2m).
\end{cases}
\end{align*}
Combining ({eq3}) and ({eq2}), we have 
\begin{align*}
a'(t) + c'(t) -4a(t)b(t)c'(t)  &= 
a'(t) -\big( -4b(t)c'(t)  +16 a(t) b(t)^2 c'(t) \big)/(4b(t)) \\ &
 = 
a'(t) +a(t)b'(t)/b(t)=(qB(t))^2/(8m).
\end{align*}
Therefore,
$
a'(t)b(t) +a(t)b'(t) = (qB(t))^2 b(t)/(8m)
$ holds. This equation yields 
\begin{align}\label{10}
a(t) = \frac{1}{8mb(t)} \int_0^t (qB(s))^2 b(s) ds.
\end{align}
Furthermore, Combining (eq1), (eq2) and \eqref{10}, 
\begin{align}\label{11}
c'(t) = \frac{a(t)}{2mb(t)} = \frac{1}{16
m^2b(t)^2} \int_0^t (qB(s))^2 b(s) ds
\end{align}
holds. Using this equation and (eq1), we also have 
\begin{align*}
b'(t) +  \frac{1}{4m^2} \int_0^t (qB(s))^2 b(s) ds = \frac{1}{2m}.
\end{align*}
This equation yields the following Hill's equation 
\begin{align*}
b''(t) + \left(
\frac{h(t)}{2m}
\right)^2 b(t) =0 , \quad b(0)=0 , \ b'(0)= \frac{1}{2m}.
\end{align*}
Thus, noting \eqref{4}, $b(t)$ can be written as
\begin{align*}
b(t) = \zeta _2(t)/(2m). 
\end{align*}
By using this, we obtain 
\begin{align*}
a(t) &= \frac{1}{8m \zeta _2(t)} \int_0^t (qB(s))^2 \zeta _2(s) ds =
 \frac{m}{2\zeta _2 (t)} \int_0^t \left(
\frac{qB(s)}{2m}
\right) ^2 \zeta _2(s)ds = \frac{-m}{2\zeta _2 (t)} \int_0^t \zeta _2''(s)
 ds \\ &=
\frac{m}{2} \left(
\frac{1 - \zeta _2'(t)}{\zeta _2(t)}
\right).
\end{align*}
Here we note that the  Wronskian of solutions of Hill's equation
\eqref{4} is constant, in particular, 
$$
\zeta _1(t) \zeta_2'(t) - \zeta _1'(t) \zeta _2(t) =1
$$
holds for all $t \in \bfR$. Then, \eqref{11} can be denoted by 
\begin{align*}
c'(t) = \frac{m }{2} \left(
\frac{1 - \zeta'_2(t)}{\zeta _2(t)^2}
\right)
 =  \frac{m }{2}
\left(
-\left(
\frac{\zeta _1(t)}{\zeta _2(t)}
\right)' +  \left(
\frac{1}{\zeta _2(t)}
\right)'
\right).
\end{align*}
Therefore  
\begin{align*}
c(t) = \frac{m}{2} \left(
\frac{1 - \zeta _1(t)}{\zeta_2(t)} 
\right) - \frac{m}{2} \left(
\frac{1 - \zeta _1(0)}{\zeta_2(0)} 
\right)
\end{align*}
holds. Here we put 
\begin{align*}
\zeta _1(t) = \rho (t)\COS{ \eta (t)}, \quad 
\zeta _2(t) = \rho (t)\SIN{ \eta (t)}, \quad 
\rho (0) = 1,\  \rho >0 ,\  \eta (0) =0,
\end{align*}
then it can be deduced that  
\begin{align*}
\rho '' (t) - \rho(t)^{-3} + \left(\frac{qB(t)}{2m}
\right)^2
\rho (t) = 0 , \quad 
\eta (t) = \int_0^t \rho (s)^{-2} ds,
\end{align*} 
see e.g. \cite{MW}. Therefore, 
\begin{align*}
\frac{1 - \zeta _1(0)}{\zeta_2(0)} = \frac{\sin ^2 (\eta (0)
 /2)}{\COS{\eta (0)/2} \SIN{\eta (0) /2}} = \TAN{\eta (0)/2} =0
\end{align*}
holds. This implies  
\begin{align*}
c(t) = \frac{m}{2} \left(
\frac{1 - \zeta _1(t)}{\zeta_2(t)} 
\right).
\end{align*}

By proving the following lemma, we have $\CAL{K} (t)$ is a propagator for
$H_0(t)$ and hence we denote $\CAL{K} (t)$ by $U_0 (t,0)$ after this Lemma. 
\begin{Lem} Let $\CAL{K}(t) ={-ia(t) x^2} e^{-ib(t)p^2} e^{i\Omega (t)L} e^{-ic(t)x^2}$. Then, for every $t \in {\bf R}$,
 $\CAL{K}(t)$ satisfies the 
{\em domain invariant property}; 
\begin{align*}
\CAL{K}(t) \D{p^2 + x^2} \subset \D{p^2 +x^2}.
\end{align*}
Moreover, for all $\phi \in L^2(\bfR ^2)$ and $t \in \bfR$, 
\begin{align} \label{mod1}
\lim_{\ep \to 0} \left\| 
\CAL{K}(t+\ep) - \CAL{K}(t)\phi 
\right\|_{L^2(\bfR ^2)} = 0
\end{align}
holds.
\end{Lem}
\Proof{ 
First, we shall prove the domain invariant property. It can be calculated that 
\begin{align*}
\left\| x^2 \CAL{K}(t)u \right\| _{L^2(\bfR ^2)}= 
\left\| 
(x+b(t)p)^2 e^{-ic(t)x^2}u 
\right\| _{L^2(\bfR ^2)}
=\left\| 
((1-2b(t)c(t))x+b(t)p)^2 u \right\| _{L^2(\bfR ^2)} .
\end{align*}
Since $\zeta _1(t)$ and $\zeta _2(t)$ are continuous, we have that both terms 
\begin{align*}
b(t) = \zeta _2 (t)/(2m), \quad 1-2b(t)c(t) = 1-(1- \zeta _1(t))/2
\end{align*}
are bounded for any fixed $t$, and hence we have $  \CAL{K}(t) \D{p^2+x^2} \subset \D{x^2}$. Next, we shall consider $p^2 \CAL{K}(t)$. By 
\begin{align*}
\left\| p^2 \CAL{K}(t) u \right\| _{L^2(\bfR ^2)} &= \left\| (p-2a(t)x)^2 e^{-ib(t)p^2} e^{-ic(t)x^2} u \right\|  _{L^2(\bfR ^2)} \\ &= 
\left\| 
 (p-2a(t)(x+2b(t)p))^2 e^{-ib(t)p^2} e^{-ic(t)x^2} u 
\right\|_{L^2(\bfR ^2)} \\ & =
\left\| 
(1-4a(t)b(t))(p-2c(t)x) -2a(t) x)^2 u
\right\| _{L^2(\bfR ^2)} \\ &= \left\| 
((1-4a(t)b(t))p + (-2a(t) -2c(t) +8a(t)b(t)c(t)))^2 u
\right\| _{L^2(\bfR ^2)}
\end{align*}
and that 
\begin{align*}
1-4a(t)b(t) &= \zeta _2'(t), \\ 
-2a(t) -2c(t) +8a(t)b(t)c(t) &= \frac{m}{\zeta _2 (t)} \left( 
\zeta _1(t)\zeta _2 '(t) -1
\right) = \frac{m\zeta _1'(t) \zeta _2 (t)}{\zeta _2 (t)} = m \zeta _1'(t),
\end{align*}
we have $\CAL{K}(t) \D{p^2+x^2} \subset \D{p^2}$ by using $\zeta _1'(t)$ and $\zeta _2'(t)$ are bounded for every fixed $t$, where we use 
$\zeta _1 \zeta _2' - \zeta _1'
\zeta _2 =1$. 
Consequantly, we have $\CAL{K}(t) \D{p^2+x^2} \subset \D{p^2+x^2}$.

At last, we prove \eqref{mod1}. For all $\phi \in C_0^{\infty}(\bfR ^2)$, denote $u(t) = \left\| 
\CAL{K}(t) \phi
\right\|_{L^2(\bfR ^2)}$. 
Clearly, we can see that the $u(t)$ is a continuous function on $\bfR_t \backslash
\CAL{A} (t)$ with    
\begin{align*}
\CAL{A} (t
)= \{ 
t \in \bfR; \zeta _2(t) =0
\}.
\end{align*}
Hence it is enough to prove $u(t)$ is a continuous function on the
neighborhood of $t \in \CAL{A}
(t)$ in the followings. Noting that $\zeta _1 \zeta _2' - \zeta _1'
\zeta _2 =1$ and $\zeta _1(t)$ is a continuous function in $t$, $\zeta
_1 (t_0)$ and $\zeta _2'(t_0)$ are not to be $0$ if $t_0 \in \CAL{A}(t)$.
Then, we have that, for all $\ep >0$ and
$t_0 \in \CAL{A}(t)$, there exists $\delta _0 >0$ such that 
\begin{align}\label{14}
\sup_{t \in [t_0- \ep , t_0 + \ep]}|\zeta _1(t)|^{-1} < \delta _0
\end{align}
holds.
By \eqref{3}, for all
$\phi \in C_0^{\infty}(\bfR ^2)$, 
\begin{align*}
e^{-ia(t)x^2}e^{-ib(t)p^2} e^{i \Omega (t) L} e^{-ic(t)x^2}\phi (x) = 
e^{-ia(t)x^2} \frac{1}{4i\pi b(t)}\int 
e^{i(x-y)^2/(4b(t))}e^{-i c(t)y^2}\phi(\hat{y}(t))dy
\end{align*}
holds, where we denote 
\begin{align*}
\hat{y}(t) := \MAT{\COS{\Omega (t)} & \SIN{\Omega (t)} \\ - \SIN{\Omega
 (t)} & \COS{\Omega (t)}} y = \hat{R}(\Omega (t)) y.
\end{align*}
Noting that 
$
{1}/{(4b(t))} -c(t) = {m \zeta _1(t)}/{(2 \zeta _2(t))}
$,  
\begin{align*}
- \frac{x \cdot y}{2b(t)}+ \left(
\frac{1}{4b(t)} -c(t)
\right)y^2 = \frac{m \zeta _1(t)}{2 \zeta _2(t)} \left(
y- \frac{x}{\zeta _1(t)}
\right)^2 - \frac{mx^2}{2 \zeta _1(t) \zeta _2(t)}
\end{align*}
holds. By the above equation and \eqref{14}, we have
\begin{align*}
&e^{-ia(t)x^2}e^{-ib(t)p^2} e^{i \Omega (t) L} e^{-ic(t)x^2}\phi (x) 
=\frac{m}{2i \pi \zeta _2(t)}  e^{i g_1(t)x^2}
\int e^{im\zeta _1(t)(y-(x/\zeta _1(t)))^2/(2 \zeta _2(t))} \phi(\hat{y}
 (t))dy \\ 
&= \frac{m}{2i \pi \zeta _2(t)\zeta _1(t)^2}  e^{i g_1(t)x^2}
\int e^{im(x-z)^2/(2 \zeta _1(t)\zeta _2(t))} \phi(\hat{z} (t)/\zeta _1(t))dz,
\end{align*}
where 
$$
g_1(t) = - \left(
a(t)- \frac{1}{4b(t)} + \frac{m}{2 \zeta _1(t)\zeta _2(t)}
\right) = \frac{-m}{2 \zeta _1(t)\zeta _2(t)}(1- \zeta _1(t)\zeta
_2'(t))= 
\frac{m\zeta _1'(t)}{2 \zeta _1(t)}.
$$
Consequently, we have 
\begin{align*}
e^{-ia(t)x^2}e^{-ib(t)p^2}e^{i \Omega (t) L}  e^{-ic(t)x^2}\phi (x) 
=
\zeta _1(t)^{-1}e^{im \zeta _1'(t)x^2/(2 \zeta _1(t))}e^{\zeta
 _1(t)\zeta _2(t)p^2/(2m)} \phi(\hat{x} (t)/\zeta _1(t))
\end{align*}
and $L^2-$ norm of the r.h.s. of above equation is clearly continuous at the neighborhood of
$\CAL{A}(t)$ by \eqref{14}.
}
\section{Hill's equation}

Now we further assume the intense of the magnetic fields is periodic function be such
that $B(t+T) =B(t)$. Then we can prove that, 
for all $t \in [NT , NT + t_0 )$, $N \in \bfN$ and $t_0 \in [0,T)$, 
\begin{align*}
U_0(t,0) = U_0(t,NT)U_0(NT,0) = U_0(t-NT,0)\big( U_0(T,0)\big)^N
\end{align*}
holds. In this section, we rewrite the $\zeta _j (t)$, $j
\in \{1,2\}$ as concrete forms thorough the $\zeta _1 (T)$, $\zeta _2(T)$ and so on. We firstly prove the following Lemma
: 
\begin{Lem}\label{L1}
For some $\theta _3 \in {\bf R}$ and $\theta _4 \in {\bf R}$, denote $\theta _1(t) = -t \theta
 _3$ and $\theta _2 (t) = - \theta _4 e^{-4t \theta _3} /(4 \theta
 _3) + \theta _4 /(4 \theta _3)$. Then 
\begin{align*}
e^{i \theta _1 (t) A} e^{-i \theta _2(t) p^2 } = e^{-it (\theta _3 A +
 \theta _4 p^2)}
\end{align*}
holds, where $A=x \cdot p + p \cdot x$.
\end{Lem}
\Proof{
Since the operator $\theta _3 A + \theta _4 p^2$ is a selfadjoint operator, we only find $\theta _1 (t)$ and
$\theta _2(t)$ such that 
\begin{align}\label{z1}
i \frac{d}{dt} e^{i \theta _1 (t) A} e^{-i \theta _2(t) p^2 } = 
\left(\theta _3 A + \theta _4 p^2  \right) e^{i \theta _1 (t) A} e^{-i \theta _2(t) p^2 }
\end{align}
holds. Then we can conclude that Lemma \ref{L1}
holds by Stone's theorem. On the other hand, by using the equation 
\begin{align*}
e^{i \theta _1(t) A}p e^{-i \theta _1(t) A} = e^{-2 \theta _1(t)} p, 
\end{align*}
\eqref{z1} can be proven easily.
}
By \eqref{3} and \eqref{301}, $U_0(T,0)$ is denoted by
\begin{align*}
U_0(T,0) = e^{-ia (T) x^2} e^{-i b(T) p^2} e^{i \Omega (T) L} e^{ia(T)
 x^2}, \quad 
U_0 (NT,0) = e^{-ia (T) x^2} e^{-i Nb(T) p^2} e^{i N\Omega (T) L} e^{ia(T) x^2}
\end{align*}
in the case where $D=\zeta _1 (T) + \zeta '_2(T) = 2$. Hence our first aim
of this section is to extend this argument for the case where $D^2 \geq 4$. 
\subsection{Proof of Lemma \ref{L50}}
At first, we consider the case where $D  \geq 2$. For all $\phi(x) \in L^2(\bfR ^2 _x)$, 
\begin{align*}
U_0(T,0)\phi(x) = \frac{1}{4 \pi i b(T)} \int e^{-ia(T) x^2}
 e^{i(x-y)^2/(4b(T))} e^{-ic(T) y^2} \phi(\hat{
y} ) dy, \quad \hat{y } = R(\Omega (T))y 
\end{align*}
holds. Here, denote that 
\begin{align*}
\tilde{U}_0(T,0) = e^{-ia_0 x^2}e^{-ia_1 A} e^{-ia_2 p^2} e^{ia_0 x^2}
 e^{i \Omega (T)L}.
\end{align*}
Then, by using the equation $e^{-ia _1A} \phi(x) = e^{-2a_1} \phi
(e^{-2a_1 } x)$ in dimension $n=2$, we have 
\begin{align*}
\tilde{U}_0(T,0)\phi(x) = \frac{e^{-2 a_1}}{4 i \pi a_2} \int 
e^{-ia_0 x^2} e^{i(e^{-2a_1}x-y)^2 /(4a_2)} e^{ia_0 y^2} \phi(\hat{y})
 dy. 
\end{align*}
Here, if the equation $U_0 (T,0) \phi = \tilde{U}_0 (T,0) \phi $ holds,
then we can see  
\begin{align*}
\frac{e^{-2a_1}}{a_2} = \frac{1}{b(T)}, \quad 
-a_0 + \frac{e^{-4a_1}}{4a_2} = -a(T) + \frac{1}{4b(T)}, \quad 
a_0 + \frac{1}{4a_2} = \frac{1}{4b(T)} -c(T).
\end{align*}
holds. Denoting $a_1 = (-1/2) \log a_3$, $a_3 >1$, we have $a_2 = b(T)
a_3$, $a_0 = a_3/(4b(T) ) + a(T) -1/(4b(T))$ and 
$$
a_3^2 -2 (1- 2 (a(T) + c(T)) b(T))a_3 + 1= a_3^2 -Da_3 +1 = 0,
$$
where we use that $2 (a(T) + c(T)) b(T) = (2-\zeta _1(T) - \zeta _2'(T))/2 = (1-D/2)$ by \eqref{301}. 
Here we put 
$$ 
\Xi _1 = \frac{2b(T)a_3 \log a_3}{a_3^2 -1}, \quad \Xi _2 = \frac{\log a_3}{2}. 
$$ 
Then we have  
$$
e^{i(\log a_3) A /2} e^{-ib(T) a_3 p^2} = e^{-i (\Xi_1 p^2 - \Xi_2 A)}
$$ 
by using Lemma \ref{L1} for $t=1$. 
By the simple calculation, we get 
\begin{align*}
\Xi_1 p^2 - \Xi_2 A &= \Xi _1 \left( p-\frac{\Xi _2}{\Xi _1} x \right)^2 - \frac{\Xi_2 ^2}{\Xi _1} x^2  \\ 
&= e^{ix^2/(2 \Xi_1)} \left( 
\Xi_1 p^2 - \frac{\Xi_2^2}{\Xi_1} x^2
\right)  e^{-ix^2/(2 \Xi_1)},
\end{align*}
 and that yields  
\begin{align*}
& e^{i(\log a_3) A /2} e^{-ib(T) a_3 p^2} = e^{-i (2b(T)a_3(\log
 a_3)p^2/(a_3^2 -1) -(\log a_3)A /2)} 
\\ &=  e^{-i(2b(T)a_3 (\log a_3)/(a_3^2 -1)) ((p -(a_3^2 -1)x/(4b(T)
 a_3))^2 -((a_3^2-1)/(4a_3b(T)))^2 x^2)} \\ 
&= 
e^{i((a_3^2 -1)/(8a_3b(T)))x^2} e^{-i(2b(T)a_3 (\log a_3)/(a_3^2
 -1))(p^2 -((a_3^2-1)/(4a_3b(T)))^2 x^2 )} e^{-i((a_3^2 -1)/(8a_3b(T)))x^2}.
\end{align*}
Therefore, Lemma \ref{L50} for the case where $D \geq 2$ is proven. 

Now we consider the case $D \leq -2$. Denote  
\begin{align*}
\Tilde{\tilde{U}}_0 (T,0) = 
(-1) e^{-ib_0 x^2} e^{-ib_1 A} e^{-ib_2 p^2} e^{ib_0 x^2} e^{i(\Omega (T)
 + \pi) L},
\end{align*}
where $(-1)$ stands for the antiperiodic part. By the same
calculation of \eqref{101} and \eqref{111}, we have $e^{i \pi L} \phi(x) =
\phi(-x)$. Paying attention to this, we also have 
\begin{align*}
\Tilde{\tilde{U}}_0(T,0)\phi(x) = \frac{-e^{-2 b_1}}{4 i \pi b_2} \int 
e^{-ib_0 x^2} e^{i(e^{-2b_1}x+y)^2 /(4b_2)} e^{ib_0 y^2} \phi(\hat{y})
 dy. 
\end{align*}
By the same way in the case where $D \geq 2$, 
\begin{align*}
-\frac{e^{-2b_1}}{b_2} = \frac{1}{b(T)}, \quad 
-b_0 + \frac{e^{-4b_1}}{4b_2} = -a(T) + \frac{1}{4b(T)}, \quad 
b_0 + \frac{1}{4b_2} = \frac{1}{4b(T)} -c(T)
\end{align*}
hold. By denoting $b_1 = (-1/2) \log b_3$, $b_3 >0$, and using the above
equations, we have $b_2 = -b(T) b_3$, $b_0 = -b_3/(4b(T)) +a(T)
-1/(4b(T))$ and 
\begin{align*}
b_3 ^2 + Db_3 + 1 =0, \quad b_3>1
\end{align*} 
hold. By using the same way for the case where $D \geq 2$, we obtain the Lemma \ref{L50}.

\subsection{Hill's equation}
Denoting  
\begin{align*}
W_0 = (B_D/T) (p^2 - C_D^2 x^2 + D_D L),
\end{align*} 
$U_0(T,0)$ can be written as
\begin{align*}
U_0 (T,0) = (-1)^{\sigma _D} e^{-iA_D x^2} e^{-iT W_0}e^{iA_D x^2},
 \quad 
U_0(NT,0) =  (-1)^{\sigma _D N} e^{-iA_D x^2} e^{-iTN W_0}e^{iA_D x^2},
\end{align*}
where $A_D$, $B_D$, $C_D$ and $D_D$ are defined in \eqref{999}.
Now we consider the asymptotic behavior of the charged particle
$e^{-it{W}_0}\phi_0$, $\phi _0 \in L^2(\bfR ^2
)$ as $t \to \infty$. Let us define the position and the momentum of
this particle at the time $t$ as
\begin{align*}
\MAT{x_w(t) \\ p_w(t)} \equiv e^{itW_0} \MAT{x \\ p}e^{-itW_0}, \quad 
\MAT{\tilde{x}_w (t) \\ \tilde{p}_w(t)} \equiv
e^{it (B_D D_D/T)L } \MAT{x_w (t) \\ p _w (t)} e^{-it (B_DD_D/T)L}
.
\end{align*} 
Then straightforward calculation shows 
\begin{align}\label{110}
\MAT{\tilde{x}''_w(t) \\ \tilde{x}''_w(t)} =( 2B_D C_D /T)^2 \MAT{\tilde{x}_w(t) \\ \tilde{p}_w(t)}
\end{align}
and 
\begin{align} \nn 
&\begin{cases}
&x_{L,1}'(t) -(B_DD_D/T) x_{L,2} (t) = 0, \\
&x_{L,2}'(t) +(B_DD_D/T) x_{L,1} (t) = 0,
\end{cases} \quad 
\MAT{x_{L,1} (t) \\ x_{L,2}(t)} :=  e^{-it (B_DD_D/T)L}
 \MAT{x_1 \\ x_2}  e^{it (B_DD_D/T)L}, \\ 
\label{101}
&\begin{cases}
&p_{L,1}'(t) -(B_DD_D/T) p_{L,2} (t) = 0, \\
&p_{L,2}'(t) +(B_DD_D/T) p_{L,1} (t) = 0,
\end{cases} \quad 
\MAT{p_{L,1} (t)  \\ p_{L,2} (t) }:=  e^{-it (B_DD_D/T)L}
 \MAT{p_1 \\ p_2}  e^{it (B_DD_D/T)L}
.
\end{align}
Here for $x_L(t) ={}^{T}(x_{L,1} (t), x_{L,2} (t)) $ and  $p_L(t) ={}^{T}(p_{L,1} (t), p_{L,2} (t)) $ \eqref{101} yields 
\begin{align} \label{111}
\begin{cases}
&x_{L} (t) = \hat{R} (B_DD_Dt/T) x \\ 
&p_{L} (t) = \hat{R} (B_DD_Dt/T) p, 
\end{cases}, \quad 
\hat{R}(t) :=  \MAT{ \COS{t} & \SIN{t} \\ - \SIN{t} & \COS{t}} .
\end{align}
Define 
\begin{align*}
\CAL{D} = 2B_D C_D/T,
\end{align*}
and then $x_w(t)$ and $p_w(t)$ can be calculated explicitly by using \eqref{110} and \eqref{111} ;
\begin{align}\label{201}
x_w(t) &= x_L (t) \cosh (t \CAL{D}) + (1/C_D)p_L(t) \sinh(t
 {\CAL{D} }), \\ 
\label{202}
p_w(t) &= p_L(t) \cosh (t {\CAL{D}}) + C_D x_L(t) \sinh (t \CAL{D}).
\end{align}
Here, the condition $D^2 >4$ appears naturally by the argument without using $W_0$; 
Let $\tilde{x}(t) = \tilde{U}_0(t,0)^{\ast} x \tilde{U}_0(t,0)$ and $\tilde{p}(t) = \tilde{U}_0(t,0)^{\ast} p
 \tilde{U}_0(t,0)$ with $\tilde{U} (t,0) = e^{-i \Omega (t) L} U_0(t,0)$, then, $\tilde{x}'(t) = \tilde{p}(t)/m$ and $\tilde{p}'(t)/m = -(qB(t)/(2m))^2 \tilde{x}(t)$
hold and hence we have Hill's equation 
\begin{align*}
\tilde{x}''(t) + \left(
\frac{qB(t)}{2m}
\right)^2 \tilde{x}(t) = 0 , \quad \tilde{x}(0)= x ,\  \tilde{x}'(0) = p/m
\end{align*}
 and differential equation $
\tilde{p}(t) = m \tilde{x}'(t) $.
Using fundamental solutions $\zeta _1$ and $\zeta_2$, we have 
\begin{align}\label{203}
\MAT{\tilde{x}(t) \\ \tilde{p}(t)} = \MAT{
\zeta _1(t) & \zeta _2(t)/m \\ m \zeta _1'(t) & \zeta _2'(t) 
} \MAT{x \\ p}.
\end{align}
Thus, by putting
$$
\CAL{A} =  \MAT{
\zeta _1(T) & \zeta _2(T)/m \\ m \zeta _1'(T) & \zeta _2'(T) 
},
$$
we have 
\begin{align*}
U_0(NT , 0)^{\ast} \MAT{x \\ p} U_0(NT ,0) =\CAL{A} ^N
 \MAT{\hat{R}(\Omega (NT)) x \\ \hat{R} (\Omega (NT))p}.
\end{align*}
and hence we can see that the asymptotic behavior of the particle
$U(NT,0) \phi$ is determined by the absolute value of eigenvalues of $\CAL{A}$ . Saying concretely, let $\lambda_0 \in
\bfR $ is the solution of 
$$
\mathrm{det}(\CAL{A} - \lambda_0) = \lambda_0 ^2 - ( \zeta _1 (T) + \zeta
_2'(T))\lambda_0 +1 = 0.
$$
Here we use $\zeta _1 \zeta _2' - \zeta _1' \zeta _2 =1$. Then, the
charged particle has only bound state if and only if $(\zeta _1(T) + \zeta _2'(T))^2 < 4$, 
the charged particle act like linearly uniform motion if and
only if $(\zeta _1(T) + \zeta _2'(T))^2 = 4$ and the charged particle
is in exponentially scattering state if and only if $(\zeta _1(T) +
\zeta_2'(T))^2 > 4$.

\subsection{Solutions of Hill's equation}

We calculate the explicit forms of fundamental solutions by using
$a(T)$, $b(T)$, $c(T)$, $\zeta_j (t-NT)$ and $\zeta '_j
(t-nT)$ in this subsection. Let us define 
\begin{align}
\MAT{x(t) \\ p(t)} = U_0(t,0)^{\ast} \MAT{x \\ p} U_0(t,0)
\end{align}
with $t \in [NT , (N+1)T]$. Here, we see that 
$U_0(t,0) = U_0(t-NT,0)U_0(NT)$ and 
\begin{align}
\MAT{x(t-NT) \\ p (t-NT)} = 
\MAT{\zeta _1(t-NT) & \zeta _2(t-NT)/m \\ 
m\zeta _1'(t-NT) & \zeta _2'(t-NT)}
\MAT{\hat{R}(\Omega (t-NT))x \\  \hat{R} (\Omega (t-NT))p}.
\end{align}
Now we calculate 
$$
\MAT{x_N \\ p_N} = U_0(NT,0)^{\ast} \MAT{x \\
p} U_0(NT,0).
$$
explicitly. We know that $U_0(NT,0) =
e^{-iA_Dx^2}e^{-iNT{W}_0}e^{iA_Dx^2}$. Thus, by putting 
$ \cosh (NT \CAL{D}) =C_N$ and $ \sinh (NT \CAL{D}) =S_N$, 
\begin{align*}
\MAT{x_N \\ p_N} &= 
(-1)^{\sigma_D N}e^{-iA_Dx^2}e^{iN T{W}_0} \MAT{\hat{x} \\ \hat{p}-2A_D \hat{x}} e^{iNT {W}_0}e^{iA_Dx^2}
\\ &= (-1)^{\sigma_D N}
e^{-iA_Dx^2} \MAT{
\hat{x} C_N + (S_N /C_D) \hat{p} \\ 
(C_DS_N -2A_DC_N) \hat{x} +(C_N
 - 2A_DS_N/C_D) \hat{p}
}e^{iA_Dx^2} \\
& \equiv 
(-1)^{\sigma_D N} \MAT{A_{1,N} & A_{2,N} \\ A_{3,N} & A_{4,N}} \MAT{ \hat{x} \\ \hat{p}}
\end{align*}
hold by \eqref{201} and \eqref{202}, where $\hat{x}= \hat{R} (B_DD_D N) x$, $\hat{p} = \hat{R} (B_DD_DN) p$ and  
\begin{align}
\nn A_{1,N} &= \cosh (NT \CAL{D}) + (2A_D/C_D) \sinh  (NT \CAL{D}) \\
\nn A_{2,N} &= (1/C_D) \sinh  (NT \CAL{D}) \\
\nn A_{3,N} &= C_D \sinh  (NT \CAL{D}) -(4A_D ^2/C_D) \sinh  (NT \CAL{D})
\\
A_{4,N} &= \cosh (NT \CAL{D}) - (2A_D/C_D) \sinh (NT \CAL{D}). \label{21}
\end{align}
Thus, we obtain
\begin{align*}
 \MAT{x(t) \\ p(t)} & \nn = 
U_0(NT,0)^{\ast} \MAT{ \zeta _1(t-NT) & \zeta _2(t-NT)/m \\ m \zeta
 '_1(t-NT) & \zeta _2'(t-NT)} \MAT{\hat{R} (\Omega (t-NT))x \\  \hat{R} (
 \Omega (t-NT))p} U_0(NT,0) \\ &= (-1)^{\sigma _D N}
\nn \MAT{ \zeta _1(t-NT) & \zeta _2(t-NT)/m \\ m \zeta
 '_1(t-NT) & \zeta _2'(t-NT)}
 \MAT{A_{1,N} & A_{2,N} \\ A_{3,N} & A_{4,N }}
 \MAT{\hat{R} (\Omega (t-NT) )\hat{x} \\
 \hat{R} (\Omega (t-NT)) \hat{p}} \\ & \equiv  (-1)^{\sigma _D N}
\MAT{B_{1,N}(t) & B_{2,N}(t) \\ B_{3,N}(t) & B_{4,N}(t)} \MAT{\hat{R }
 (\Omega (t+N \sigma_D \pi))x \\ \hat{R} (\Omega (t + N \sigma_D \pi))p} \\ &= 
\MAT{B_{1,N}(t) & B_{2,N}(t) \\ B_{3,N}(t) & B_{4,N}(t)}
 \MAT{\hat{R }
 (\Omega (t)x \\ \hat{R} (\Omega (t)p}
, 
\end{align*}
where  
\begin{align*}
\nn B_{1,N}(t) &= A_{1,N}\zeta _1(t-NT) +(A_{3,N}/m)\zeta _2(t-NT), \\
\nn B_{2,N}(t) &= A_{2,N}\zeta
 _1(t-NT) + (A_{4,N}/m)\zeta _2(t-NT), \\ 
\nn B_{3,N}(t) &= mA_{1,N} \zeta _1'(t-NT) + 
 A_{3,N} \zeta _2'(t-NT), \\ 
B_{4,N}(t) &= mA_{2,N}\zeta _1'(t-NT) + A_{4,N}\zeta
 _2'(t-NT). 
\end{align*}
On the other hand, we see that 
$$
\MAT{x(t) \\ p(t)} = \MAT{\zeta _1(t) & \zeta _2(t)/m \\ m \zeta '_1(t) &
\zeta _2'(t)} \MAT{\hat{R} (\Omega (t))x \\ \hat{R} (\Omega (t))p},
$$
hold and which yields
\begin{align}\label{32}
\begin{cases}
\zeta _1(t) &= B_{1,N}(t) ,\\
\zeta'_1(t) &= B_{3,N}(t)/m ,
\end{cases}
 \quad 
\begin{cases}
\zeta _2(t) &= m B_{2,N}(t), \\ 
\zeta'_2(t) &= B_{4,N}(t),
\end{cases}
\end{align}
for all $t \in [NT, (N+1)T)$. Indeed, $
 \zeta _1(t)\zeta _2'(t) - \zeta _1'(t)\zeta _2(t)=1$ holds by
\begin{align}
A_{1,N} A_{4,N} - A_{2,N}A_{3,N} = \cosh ^2(NT \CAL{D}) - \sinh ^2 (NT
 \CAL{D}) =1.\label{22}
\end{align}

We now rewrite the condition of the magnetic fields of \cite{Ko} by
using $D$, $\zeta_j(T)$ and $\zeta_j'(T)$. If $\zeta _1 (t) = e^{- \lambda t} \chi_1(t)$, $\lambda >0$ holds, it
must be proven that $A_{1,N} = A_{3,N} = \CAL{O}({e^{- \lambda N}})$. 
Suppose that $\zeta _2(T) \neq 0$, which yields $(1/C_D) \neq  \infty$. 
Here paying attention to 
\begin{align*}
NT \CAL{D} = 2NC_DB_D = 
\begin{cases}
N \LOG{a_3} > 0, & D>2, \\ 
-N \LOG{b_3} <0, & D< -2,
\end{cases}
\end{align*}
we have 
\begin{align*}
A_{1,N } &= \begin{cases}
a_3^N (1+ 2A_D/C_D)/2 + \ord{a_3}{-N}, & D>2, \\ 
b_3^N (1- 2A_D/C_D)/2 + \ord{b_3}{-N},& D < -2, 
\end{cases} 
\\ &= 
  \begin{cases}
a_3^{N} (a_3 (a_3 - \zeta _2'(T)) /(a_3^2 -1)) +  \ord{a_3}{-N}, & D>2, \\ 
b_3^{N} (b_3(b_3+ \zeta _2'(T)) /(b_3^2 -1)) +  \ord{b_3}{-N},  & D < -2, 
\end{cases}.
\end{align*}
Thus, we have $A_{1,N} < C e^{ - \lambda  N}$ with $\lambda >0$ if
$\zeta _2'(T) =  a_3$ or
$\zeta _2'(T) = -b_3$ holds.
\begin{Rem}
In the case where $D =2$, 
\begin{align*}
(1/ C_D) \sinh (NT \CAL{D}) &= (2a_3 b(T)(a_3^N +1)/(a_3^N
 (a_3+1)))(a_3^{N-1} + a_3^{N-2} +... + a_3+1) \\ & \to 
 2N b(T) \quad \mathrm{as} \quad a_3 \to 1 
\end{align*}
holds. The case where $D  = -2$ can be calculated by the same way. 
\end{Rem}

\section{Relative compactness of resolvent of Floquet Hamiltonian}
Now we prove so-called relative compactness of weighted
resolvent. Let us define $X_f(z) = f(\hat{H}_0 -z) f$ with $(f
\phi)(t,x) = f(|x|)\phi(t,x)$, $\phi \in \SCR{K}$ and $z \in
\bfC \backslash \bfR$ where $f \in C_0^{\infty}(\bfR)$ satisfies 
$f(s) = 1 $ if $s \leq R_0$ and $f(s) = 0$ if $s > R_0 +1$ for some $R_0 >0$.
Here we prove that $X_f(z)$ is a compact operator on
$\SCR{K}$ for all $\Im z \in \bfR \backslash \{0\}$ and
$\Re z \in \bfR$. By the same argument of Yajima \cite{Ya}, 
\begin{align}\nn
(X_f(z)\phi)(t,x) &= i f(|x|) \sum_{N=1}^{\infty} \int_0^T e^{i(t+NT -s)z}
 U_0(t+NT ,s)(f \phi)(s)ds  \\ & \quad + i f(|x|) \int_0^t e^{i(t-s)z} U_0(t,s)(f
 \phi)(s) ds \label{31}
\end{align}
holds. Here we supposed that $\Im z >0$. For the case of $\Im
z <0$ can be calculated by the same way. By Theorem \ref{T1},
we see that 
\begin{align*}
&fU_0({\tau},s)(f \phi)(s,x)  \\ &=
 e^{-ia({\tau})x^2} e^{i(\Omega (\tau) - \Omega (s))L}f(|x|)e^{-ib({\tau})p^2}e^{-i(c({\tau})-c(s))x^2} (e^{ib(s)p^2} f(|x|)
 e^{-ia(s)x^2})\phi(s,x) \\ 
&= e^{i \tilde{\Omega}L} e^{-ia({\tau})x^2}f(|x|) \frac{ \lim_{\ep \to 0}}{(4 \pi i)^2 b({\tau})b(s)} \int 
e^{i|x-z|^2/(4b({\tau}))} e^{-\ep z^2}e^{-i \tilde{c}({\tau},s)z^2} \int
 e^{-i(z-y)^2/(4b(s))} g(s,y)dydz
\end{align*}
where $\tilde{\Omega} = \Omega (\tau) - \Omega (s)$, $\tilde{c}({\tau},s) = c({\tau})-c(s)$ and $g(s,y) = e^{-ia(s)y^2}f \phi
(s,y)$. After the simple calculation, we also see that 
\begin{align*}
(fU_0({\tau},s)f \phi)({\tau},x) = e^{i \tilde{\Omega}L} e^{-ia({\tau})x^2}f(|x|) \lim_{\ep \to 0} \frac{1}{(4 \pi i)^2 b({\tau})b(s)}
 \int g(s,y)dy \int e^{g_2(x,y,z)} dz
\end{align*}
with
\begin{align*}
&g_2(x,y,z) \\ & = - \left(
\ep + \frac{i}{4}\left( \frac{1}{b(s)} - \frac{1}{b({\tau})}\right) + i \tilde{c}({\tau},s)
\right)z^2 - \frac{i}{2} \left(
\frac{x}{b({\tau})} - \frac{y}{b(s)}
\right) \cdot z + \frac{i}{4} \left(
\frac{x^2}{b({\tau})} - \frac{y^2}{b(s)}
\right).
\end{align*}
Define $\Gamma _1({\tau},s) = \zeta _1(s) / \zeta _2(s) - \zeta _1({\tau}) /\zeta
_2({\tau})$. Then by the definitions of $b({\tau})$ and $c({\tau})$,
$g_2(x,y,z)$ can be calculated that 
\begin{align*}
g_2(x,y,z)  & =
- \left( \ep + \frac{im \Gamma_1({\tau},s)}{2} \right) \left(
z + \frac{i}{2(\ep + im \Gamma _1({\tau},s)/2)} \left(
\frac{mx}{\zeta _2({\tau})} - \frac{my}{\zeta _2(s)}
\right)
\right)^2 \\ & 
- \frac14 \left(\ep + \frac{im \Gamma _1({\tau},s)}{2} \right)^{-1} \left(
\frac{mx}{\zeta _2({\tau})} - \frac{my}{\zeta _2(s)}
\right)^2 + \frac{mi}{2} \left(
\frac{x^2}{\zeta _2({\tau})} - \frac{y^2}{\zeta _2(s)}
\right).
\end{align*}
Here, we put 
\begin{align*}
\Gamma _2({\tau},s,x,y) =\frac{i}{2m \Gamma _1({\tau},s)} \left(
\frac{mx}{\zeta _2({\tau})} - \frac{my}{\zeta _2(s)}
\right)^2 + \frac{mi}{2} \left(
\frac{x^2}{\zeta _2({\tau})} - \frac{y^2}{\zeta _2(s)}
\right).
\end{align*}
Then, we have 
\begin{align*}
\lim_{\ep \to 0} \int e^{g_2(x,y,z)} dz 
= \left(
\frac{2 \pi}{m |\Gamma _1({\tau},s)|}
\right) e^{\Gamma _2({\tau},s,x,y)},
\end{align*}
and we also have 
\begin{align}\nn
&(fU_0({\tau},s)f \phi )({\tau},x) \\ &= \left(
\frac{2 \pi}{m |\Gamma _1({\tau},s)|}
\right) \frac{m^2}{4( \pi i)^2 \zeta _2({\tau}) \zeta _2(s)}
 e^{i(\Omega (\tau) - \Omega (s))L} e^{-ia({\tau})x^2}f(|x|)
\int  e^{\Gamma _2({\tau},s,x,y)} (g )(s,y)dy. \label{39}
\end{align}
\begin{Thm}\label{T10} Suppose $D^2 \geq 4$.
Let $(f \phi)(t,x) = f(|x|)\phi(t,x)$, $\phi \in \SCR{K}$
 and define $X_f$ as 
\begin{align}
X_f = (f(\hat{H}_0 -z)^{-1} f)
\end{align}
for all $z \in \bfC \backslash \bfR$. Then $X_f$ is a compact operator
 on $\SCR{K}$.
\end{Thm}
\Proof{
Define 
\begin{align*}
&\Gamma _3 (\tau, s) = \zeta _2 (\tau) \zeta _2(s) |\Gamma _1 (\tau,s)|, \quad
\CAL{S}_1^N (t,s) = \left\{
(t,s) \in [0,T)^2 ; |\Gamma _3(t+NT,s)| < \ep 
\right\}, \\ & \CAL{S}_2^N (t,s )=[0,T)^2 \backslash \CAL{S}_1^N 
\end{align*}
for sufficiently small $\ep >0$. Then 
\begin{align*}
&\sum_{N=0}^{\infty} 
\left\|
f(|x|) \int_{\CAL{S}_1^N(t,s)}e^{i(t+NT -s)z}U_0(t+NT,s)(f \phi)(s, x) ds 
\right\|_{\SCR{K}} 
\\ & \leq C \sum_{N=0}^{\infty} e^{-|\Im z|NT} |\CAL{S}_1^N
 (t,s)|_{\SCR{L}}, \quad \left| 
\CAL{S}_1^N (t,s)
\right|_{\SCR{L}} = \left| 
\int_{\CAL{S}_1 (t,s) } ds dt
\right| .
\end{align*}
holds. On the other hand, define $\CAL{I}$ as
\begin{align*}
(\CAL{I} \phi)(t,x) \equiv i \sum_{N=0}^{\infty} 
\int_{\CAL{S}_2^N(t,s)} 
f(|x|)e^{i(t+NT-s)z} U_0(t+NT,s) (f \phi) (s,x) ds,
\end{align*}
and $\CAL{I} (t,x)$ as  
\begin{align*}
\CAL{I}(t,x) &:= C_M \sum_{N=0}^{\infty} 
\int_{\CAL{S}_2^N(t,s)} 
e^{i(t+NT-s)z} (\Gamma _3 (t+NT , s))^{-1}
e^{-ia(t+NT)x^2}f(|x|) \\ & \qquad \qquad \times
\int 
e^{\Gamma _2 (t+NT,s,x,y)}
e^{ia(s) y^2} f(|y|) ds dy
\end{align*}
with $C_M = -m/ (2\pi)$. Noting that $
\Gamma _3(t,s) = |\Gamma _1(t,s)| \zeta _2(t) \zeta _2(s)
$ holds, we easily see that 
\begin{align*}
\int_0^T \int \CAL{I} (t,x) dx dt < \infty 
\end{align*}
by the definition of $\CAL{S}_2^N(t,s)$, and it implies $\CAL{I}$ is
a compact operator. By these results, we have that 
\begin{align*}
(X_f(z)\phi)(t,x) = \CAL{I}(t,x)\phi(t,x) + \ep (t,x) \phi(t,x),
\end{align*}
where $\ep(t,x)$ is a bounded operator satisfies $\| \ep (t,x)\| \leq C \sum_{N=0} ^{\infty} e^{-|\Im z| NT} |\CAL{S}_{1,N}(t,s) |_{\SCR{L}}$.
By using the following Lemma \ref{L17}, we have the Theorem \ref{T10}.
}

\begin{Lem}\label{L17}
For all sufficiently small $\ep >0$, there exists sufficiently small $\delta
 >0$ such that
\begin{align}\label{30}
|\CAL{S}_1^N(t,s)|_{\SCR{L}} \leq \delta. 
\end{align}
holds where $\CAL{S}_1^N(t,s)$ and $|\CAL{S}_1^N (t,s)|_{\SCR{L}}$ are
 equivalent to those in the proof of theorem \ref{T10}.
\end{Lem}
\Proof{
We see that
\begin{align*}
\Gamma _3(t+NT,s) 
&= \zeta _2(t+NT) \zeta _1(s) - \zeta _1(t+NT) \zeta
 _2(s),
\\ &=
(mA_{2,N} \zeta _1(t) + A_{4,N} \zeta _2(t))\zeta _1(s)
-(A_{1,N} \zeta _1(t) + A_{3,N} /m )\zeta _2(s).
\end{align*}
Hence we define that 
\begin{align*}
\CAL{M}_1(t) &=
\zeta _2(t+NT) \zeta _1(s) - 
\zeta _1(t+NT) \zeta _2(s) 
= \Gamma _3(t+NT,s), \\ 
\CAL{M}_2(t) &=
\zeta _1(t+NT) \zeta _1(s) + 
\zeta _2(t+NT) \zeta _2(s) ,
\\ 
\CAL{N}_1(s) &= \zeta _2(t+NT) \zeta _1(s) - 
\zeta _1(t+NT) \zeta _2(s) =
 \Gamma _3(t+NT,s) , \\ 
\CAL{N}_2(s) &= \zeta _1(t+NT) \zeta _1(s) + 
\zeta _2(t+NT) \zeta _2(s),
\end{align*}
then simple calculation shows that $\zeta _1'(t+NT) \zeta _2(t+NT) - \zeta _1
(t+NT) \zeta _2'(t+NT) = (A_{1,N} A_{4,N} -A_{2,N} A_{3,N}) = 1$ and 
\begin{align*}
& \CAL{M}'_1(t) \CAL{M} _2(t) - \CAL{M}_1(t) \CAL{M}_2'(t) = 
 ( \zeta _1(s)^2 + \zeta
 _2(s) ^2) \\ 
& \CAL{N}'_1(s) \CAL{N} _2(s) - \CAL{N}_1(s) \CAL{N}_2'(s) = 
- (\zeta _1(t+NT)^2 + \zeta _2(t+NT)^2)
\end{align*} 
holds. By the virtue of the above equations, we have that if $\CAL{M} _1(t) = 0$,
then, $\CAL{M} _1'(t) \neq 0$ holds. That means the zero points of
$\Gamma _3(t+NT ,s)$ in $t$ has at most single multiplicity. On the same
arguments for $\CAL{N}$ show that the zero point of $\Gamma (t+NT
,s)$ in $s$ have at most single multiplicity. Thus \eqref{30} holds.
}

\section{Mourre theory}
In this section, we consider the Mourre estimate for the case where $D^2>4$. 
This discussion is based on \cite{BCHM}. Let us define 
\begin{align*}
\hat{W} = -i \partial _t + \Sigma _1 p^2 + \Sigma _2 x^2 + \Sigma _3 L +
 \Sigma _4 + J^{\ast}VJ
\end{align*}
with $\Sigma _1 \Sigma _2 < 0$. Here $L = x_1 p_2 - x_2 p_1$ and $J$ is a
unitary operator on $\SCR{K}$ and satisfies that for some bounded functions $\theta _j (t)$, $j \in \{1,2,3,4\}$, 
$$ J^{} \MAT{x \\ p} J ^{\ast}= \MAT{\theta _1 & \theta _2 \\ \theta _3 & \theta _4} \MAT{x \\ p}.
$$ 
In this section, we
suppose that $V$ satisfies Assumption \ref{A4}. For time-periodic magnetic fields, we choose $J = \SCR{J}_D$ and 
\begin{align*}
D>2 &\Rightarrow \Sigma _1 = B_D/T , \quad 
\Sigma _2 = -(B_DC_D^2)/T , \quad 
\Sigma _3 = -\Omega(T)/T, \quad \Sigma _4 =0, \\ 
D<-2 &\Rightarrow \Sigma _1 = B_D/T , \quad \Sigma _2 =  -(B_DC_D^2)/T , \quad 
\Sigma _3 = -(\Omega (T) + \pi)/T,  \quad  \Sigma_4 = \pi /T.
\end{align*}
Denote that $\hat{H} = J \hat{W} J^{\ast}$ and 
\begin{align*}
X_0 = \left(\sqrt{ - \Sigma _2 / \Sigma _1}\right) /2 , \quad X_1 = -\left(\sqrt{-\Sigma
 _1/\Sigma _2} \right)/4. 
\end{align*}
Let us define $\hat{K}$ as follows:
\begin{align*}
\hat{K} &= e^{iX_1 p^2}e^{-iX_0 x^2} \hat{W} e^{iX_0 x^2}e^{-iX_1 p^2}
= 
 -i \partial _t + 2X_0 \Sigma _1 A + \Sigma _3 L + e^{iX_1 p^2}e^{-iX_0
 x^2}J^{\ast}VJe^{iX_0 x^2}e^{-iX_1 p^2} .
\end{align*}
Here, by \cite{BCHM}, one can expect the candidate of conjugate operator
$\SCR{A} _0$ for
$\hat{K}$ is 
\begin{align*}
\SCR{A}_{0} = ( \log \J{x}- \log \J{p}  )/(8 X_0\Sigma _1).
\end{align*}
Let us write 
$\hat{K} = \hat{K}_{0} +\hat{V}$, $\hat{V} = e^{iX_1 p^2}e^{-iX_0 x^2}
 J^{\ast}VJ e^{iX_0 x^2}e^{-iX_1 p^2}
$. 
Then, paying attention to 
$[-i\partial _t , \SCR{A}_0]=[L, \SCR{A}_0] =0$, 
\begin{align*}
i[\hat{K}_{0} , \SCR{A}_0] &= i[x \cdot p + p \cdot x ,  \log \J{x} -
 \log \J{p} ]/4  = 
((p)^2\J{p}^{-2} + x^2\J{x}^{-2})/2 =
1-( \J{p}^{-2}+ \J{x}^{-2}) /2
\end{align*}
holds. For all $R \gg 1$,
$F(|p^2+x^2| \leq R)$ is a compact operator on $\SCR{K}$. Thus, there exists a compact operator $\CAL{K}_0$ such that
\begin{align*}
i[\hat{K}_{0} , \SCR{A}_0] &=1- F(|p^2+x^2| \geq R) \frac{\J{p}^{-2}+ \J{x}^{-2}}{2}F(|p^2+x^2| \geq R) 
+ \CAL{K}_0
\end{align*}
holds. By the G\aa rding inequality, for sufficiently large $R \gg 1$, 
$$
F(|p^2+x^2| \geq R)  \frac{ \J{p}^{-2}+ \J{x}^{-2}}{2}F(|p^2+x^2| \geq
R) < 3/4
$$
holds. This equation yields following Mourre estimate; Let us define $\varphi$ as in Theorem \ref{T2} and 
$$
\SCR{A}= e^{iX_0 x^2}e^{-iX_1 p^2} \SCR{A}_0 e^{iX_1p^2}e^{-iX_0 x^2} 
$$ 
Then, we have 
\begin{align*}
(i[\hat{W}_0 , \SCR{A}] \varphi (\hat{H}) \phi , \varphi (\hat{H}) \phi) &= (i[\hat{K}_0 , \SCR{A}_0]
e^{iX_1 p^2} e^{-iX_0 x^2} \varphi(\hat{H}) \phi , e^{iX_1 p^2} e^{-iX_0 x^2} \varphi (\hat{H})\phi ) \\ & \geq 
1/4  + (\CAL{K}_0e^{iX_1 p^2} e^{-iX_0 x^2} \varphi (\hat{H}) \phi , e^{iX_1 p^2} e^{-iX_0 x^2} \varphi (\hat{H})\phi) .
\end{align*}
Hence, one can see the conjugate operator of
$\hat{H}$ is 
\begin{align*}
\hat{A}= J \SCR{A} J^{\ast} \equiv ( \LOG{\J{\theta _1 x + \theta _2 p}} - \LOG{
\J{\theta _3 x + \theta _4 p}})/(8X_0 \Sigma _1),
\end{align*}
where we suppose that $\theta _j$ is defined as bounded
function. (in the case of time-periodic magnetic fields, we calculate it
in \S{4.1} and proves that these are bounded function). Indeed, putting
$\psi _{0} =\varphi(\hat{H}) \phi $, we can obtain 
\begin{align*}
(i[\hat{H}, \hat{A}]\psi _0 , \psi _0) &= (J i[\hat{W}_{0} , \SCR{A}]J^{\ast} \psi _0 , \psi _0) \\ & \quad + 
(i[V ,  ( \LOG{\J{\theta _1 x + \theta _2 p}} - \LOG{\J{\theta _3 x + \theta _4 p}})/(8X_0 \Sigma _1)]
\psi _0 , \psi _0) \\ & \geq 1/4  + (\CAL{K}_0e^{iX_1 p^2} e^{-iX_0 x^2} \psi_0 , e^{iX_1 p^2} e^{-iX_0 x^2}\psi_0 ) \\  & \quad
+  (i[V ,  ( \LOG{\J{\theta _1 x + \theta _2 p}} - \LOG{\J{\theta _3 x + \theta _4 p}})/(8X_0 \Sigma _1)]
\psi _0 , \psi _0)
\end{align*}
holds. Here we need to prove thet $\varphi (\hat{H})i[V,\LOG{\J{\theta _1 x + \theta _2 p}} \varphi (\hat{H})$
and  $\varphi (\hat{H})i[V,\LOG{\J{\theta _3 x + \theta _4 p}}]\varphi (\hat{H})$ are compact
operators. We only calculate about $i[V,\LOG{\J{\theta _1 x + \theta _2
p}}]$. Denote $h_1 \in C_0^{\infty} (\bfR)$ and $h_2, h_2^I \in L^{\infty}
(\bfR)$ as follows
\begin{align*}
h_1 (t) = 
\begin{cases}
1, & |t| \leq 1/2 ,\\ 
0, & |t| \geq 1,
\end{cases}  \quad h_2(t) = 
\begin{cases}
1, & t \leq \sqrt{2}, \\
0, & t \geq \sqrt{2},
\end{cases} \quad 
h_2^I(t) = 1 - h_2(t).
\end{align*}
Noting that
\begin{align*}
\varphi(\hat{H})i[V , \LOG{\J{\theta _1 x + \theta _2 p}}] \varphi(\hat{H})
 = 
(-1/2) \varphi(\hat{H}) i[\LOG{1+
 (\theta _1 x + \theta _2 p)^2} , V]  \varphi (\hat{H}),
\end{align*} 
we define $L_{R_0}(t) = h_1(t/R_0) \LOG{1+
 t}$ for some $R_0 > 0$, and apply the Helffer-Sj\"{o}strand formula
 (Helffer-Sj\"{o}strand \cite{HS}, see also \cite{DG}) to
 $L_{R_0}((\theta _1 x + \theta _2 p)^2)$. Then we have 
\begin{align*}
L_{R_0}((\theta _1 x + \theta _2 p)^2) = (2 \pi i)^{-1} \int
 \bar{\partial }_z l_{R_0}(z) (z- (\theta _1 x + \theta _2 p)^2) ^{-1}dz d \bar{z},
\end{align*}
where $l_{R_0}$ stands for an almost analytic extension of
$L_{R_0}$, which is defined as 
\begin{align*}
l_{R_0}(z) = \sum_{n=0}^{N-1} \frac{i^n}{n !} \left( 
\frac{d^n}{d \tau ^n} \left( 
h_1(\tau/R_0) \LOG{1+ \tau}
\right)
\right) \kappa ^n h_1 (\kappa / (2\J{\tau} )), \quad z = \tau + i \kappa.
\end{align*}
Simple calculation shows that there exist sufficiently small constant $0< \ep _2 \ll 1$ and a constant
$c_1 >0$, which is independent of $R_0$, such that 
\begin{align}\label{1005}
|\bar{\partial}_z l_{R_0} (z)| \leq c_1 \J{z}^{\ep _2 -1-M} |\Im z|^M,
 \quad M \in \bfN
\end{align}
holds. Then 
\begin{align*}
i[L_{R_0}((\theta _1 x + \theta _2 p)^2/R_0), V] &= \frac{\theta _2}{2
 \pi i} \int 
\bar{\partial }_z l_{R_0}(z) (z- (\theta _1 x + \theta _2 p)^2)^{-1} \Big(
 (\theta _1 x + \theta _2 p) \cdot \nabla V \\ & + \nabla V \cdot (\theta _1
 x + \theta _2 p) \Big) (z- (\theta _1 x + \theta _2 p)^2)^{-1} dz
 d \bar{z} 
\end{align*}
holds. Here we can obtain that for sufficiently small $0< \gamma _2 \ll 1$ 
\begin{align}\label{1004}
|\nabla V| (z- (\theta _1 x + \theta _2 p)^2 )^{-1} \J{x}^{\gamma _2} \leq 
C(|\Im z|^{-1} + |\Im z|^{-1- \gamma _2/2} (1 + \J{z}^{\gamma _2/2} ))
\end{align}
holds. Indeed, by using 
\begin{align*}
& (z- (\theta _1 x + \theta _2 p)^2)^{-1}x^2 = -2 \theta _2^2 (z- (\theta
 _1 x + \theta _2 p)^2)^{-2} + x^2  (z- (\theta
 _1 x + \theta _2 p)^2)^{-1} \\ & 
-4i \theta _2 x \cdot (\theta _1 x + \theta _2 p)  (z- (\theta
 _1 x + \theta _2 p)^2)^{-2} -8 \theta _2^2  (\theta _1 x + \theta _2
 p)^2   (z- (\theta
 _1 x + \theta _2 p)^2)^{-3} 
\end{align*} 
and 
\begin{align}\nn
\left| 
(z-t^2)^{-1} \J{t}
\right| &= \left| 
(z-t^2)^{-1} (h_2(t \J{z}^{-1/2}) + h_2^{I} (t \J{z}^{-1/2})) \J{t}
\right| \\ & \leq \sqrt{2}( \J{z}^{1/2}|\Im z|^{-1} + \J{z}^{-1/2}), \label{1006}
\end{align} 
we can get 
\begin{align} \label{mod2}
& \left\| |\nabla V|^{2/ \gamma _2}(z- (\theta _1 x + \theta _2 p)^2
 )^{-1}
\J{x}^2
 \right\|_{\SCR{K}} \leq C \left( 
|\Im z|^{-1} +  |\Im z|^{-2} (1 + \J{z} ^{})
\right) \\ 
& 
 \left\| |\nabla V|^{0}(z- (\theta _1 x + \theta _2 p)^2
 )^{-1}
\J{x}^0
 \right\|_{\SCR{K}} \leq |\Im z|^{-1}.  \label{mod3}
\end{align} 
By interpolating \eqref{mod2} and \eqref{mod3}, we have \eqref{1004}. By \eqref{1005}, \eqref{1004} and \eqref{1006}, one can obtain that 
\begin{align}\nn
& \left\| \left| \bar{\partial }_z l_{R_0}(z) \right| \left|\J{(\theta _1 x + \theta _2 p)}(z-
 (\theta _1 x + \theta _2 p)^2)^{-1}\right|\left| |\nabla V|
 (z- (\theta _1 x + \theta _2 p)^2)^{-1} \J{x}^{\gamma _2}\right| \right\|_{\SCR{K}} \\  & \leq 
 C \J{z}^{\ep_2-7/2+ \gamma _2/2} |\Im z|^{1- \gamma _2/2} \leq 
 C
 \J{z}^{-5/2 + \ep_2}
  \label{1001} 
\end{align}
for some constant $C>0$, which is independent on $R_0$. 
Noting that $\J{z}
\geq \tau \geq s/2$ on the support of $h'_1(\tau /s)$,  
\begin{align*}
\left| h_1 ( \tau /R_1) - h_1 (\tau / R_2)\right| \leq 
C \int_{R_2}^{R_1} \left| 
s^{-1-\ep_3 } h_1'(\tau /s) \right| ds 
 \J{z}^{\ep_3 }
\end{align*}
holds for some small $\ep_3 >0$. The above equation and \eqref{1001} yield  
\begin{align*}
& ((i[L_{R_1}((\theta _1 x + \theta _2 p)^2), V]-i[L_{R_2}((\theta _1
 x + \theta _2 p)^2), V] )\J{x}^{\gamma _2} \cdot \J{x}^{- \gamma _2} \psi _0 , \psi _0) \\ & \to 0, \quad \mathrm{as}\ R_1,R_2
 \to \infty, 
\end{align*}
by taking $\ep_2 + \ep_3 
\leq 1/4$, and hence we have that $i\varphi(H)[V,\LOG{\J{\theta _1 x + \theta _2 p}}] \varphi (H)$
 is a compact operator since $\J{x}^{-\gamma _2} \varphi (H)$ is a compact operator by Theorem \ref{T10}. By \eqref{1001}, one can also obtain that
 $i[[\hat{H}, \hat{A}], \hat{A} ]$ is a bounded operator under
 Assumption \ref{A4}. Other
 conditions, which is necessary for to prove the
 Mourre theory, see e.g. \cite{CFKS}, can be proven easily since
 $i[H,A]$ is a bounded operator.

\subsection{Calculation of $\theta _j$}
Now, we calculate $\theta _j$ defined above, where we suppose $D>2$. We only need to
calculate 
\begin{align*}
\MAT{x_{\theta} \\ p _{\theta}} = {\SCR{J}_D} e^{iX_0 x^2}e^{-iX_1 p^2}
 \MAT{x \\ p} e^{iX_1 p^2} e^{-i X_0 x^2} {\SCR{J}_D}^{\ast}.
\end{align*}
Denote that 
\begin{align*}
\CAL{Q}_1 (t) &= \MAT{1+4X_0X_1 & -2X_1 \\ -2X_0 & 1}, \\ 
\CAL{Q}_2 (t) &= \MAT{\cosh (t \CAL{D}) + (2A_D/C_D) \sinh (t \CAL{D}) &
 (1/C_D) \sinh (t \CAL{D})
 \\  2A_D \cosh (t \CAL{D}) + C_D \sinh (t \CAL{D}) & \cosh (t \CAL{D})}, \\ 
\CAL{Q}_3(t) &= \MAT{\zeta _2'(t) & - \zeta _2(t)/m \\ -m \zeta _1'(t) &
 \zeta _1(t)}, \quad 
\hat{R}_1(t) = \hat{R} (\Omega _1(t)), \quad \Omega_1(t) = \Omega (T)t/T -
 \Omega (t).
\end{align*}
Then, 
\begin{align}\label{y1}
\MAT{\theta _1 & \theta _2 \\ \theta _3 & \theta _4} = 
\CAL{Q}_3(t) \CAL{Q}_2(t) \CAL{Q}_1(t) \hat{R}_1(t)
\end{align}
holds. Indeed, straightforward calculation shows that 
\begin{align}\label{y2}
e^{-ia(T) x^2} e^{itW_0} \MAT{x \\ p} e^{-it W_0} e^{ia(T) x^2} =
 \CAL{Q}_2(t) \hat{R}(\Omega (T)t/T) \MAT{x \\ p}
\end{align}
holds by \eqref{201} and \eqref{202}. Moreover, if the matrix $\CAL{Q}_4(t)$ and
$\hat{R}_0(t)$ satisfy that 
\begin{align*}
U_0(t,0) \MAT{x \\ p} U_0(t,0)^{\ast} = \CAL{Q}_4(t) \hat{R}_0 (t) \MAT{x \\ p},
\end{align*} 
then, 
\begin{align}\label{y3}
\CAL{Q}_4(t) \hat{R}_0(t) U_0(t,0)^{\ast} \MAT{x \\ p} U_0(t,0) = \CAL{Q}_4 \hat{R}_0 (t) \CAL{Q}_3 (t) \hat{R} (\Omega (t)) \MAT{x \\ p} 
= \MAT{x
 \\ p}
\end{align}
holds. By using \eqref{203} and $\zeta _1 \zeta _2' - \zeta _1' \zeta _2
=1$, we have $
\CAL{Q}_4 (t)^{-1} = \CAL{Q}_3(t) ,  \hat{R}_0(t) =  \hat{R}(\Omega (t))^{-1}
$. By \eqref{y2} and \eqref{y3}, we obtain \eqref{y1}. For the case where $D<-2$ can be calculated by the same way.

\appendix

\section{Model}
Now we consider the model of the magnetic fields which satisfies
Assumption \ref{A1}. For the case where the magnetic fields can be written by $B(t) = \lambda_0 \theta (t)$
with arbitrarily time periodic function $\theta (t)$ are considered by
\cite{CS} and it is proven that there exists an
interval $\Lambda \subset (- \infty , \infty)$ such that for all $\lambda_0
\in \Lambda$, $D^2 > 4$ holds. 
However, it is difficult to calculate $\Lambda$ precisely and to
calculate an additional condition
$\zeta _2(T) \neq 0$. 

To avoid this, we consider the case where $B(t)$ is defined as the
following three pulsed magnetic field:
\begin{align*}
B(t) = 
\begin{cases}
B_1 & 0 \leq t \leq T_1 \\
B_2 & T_1 \leq t \leq T_2 \\
B_3 & T_2 \leq t \leq T
\end{cases}
, \quad 0<T_1 \leq T_2  <T, \ B(t+T) = B(t).
\end{align*}  
Then fundamental solutions of Hill's equation \eqref{4} are 
\begin{align*}
\zeta _1(t) &= 
\begin{cases}
&\COS{\omega_1 t /2}, \\ 
&\COS{\omega_1T_1/2}\COS{\omega _2 (t-T_1)/2}- (\omega _1/\omega
 _2)\SIN{\omega _1 T_1/2} \SIN{\omega_2(t-T_1)/2}, \\  
&\alpha _1 \COS{\omega _3 (t-T_2)/2} + (2 \alpha _2 / \omega _3)
 \SIN{\omega _3 (t-T_2)/2},
\end{cases}
\\ 
\zeta _2(t) &= 
\begin{cases}
&(2/\omega_1)\SIN{\omega_1 t /2}, \\ 
&(2/ \omega _1) \SIN{\omega _1 T_1 /2} \COS{\omega _2 (t-T_1)/2} +
 (2/\omega _2) \COS{\omega _1 T_1 /2} \SIN{\omega _2 (t-T_1)/2}, \\
&\beta _1 \COS{\omega _3 (t-T_2)/2} + (2 \beta _2/ \omega _3) \SIN{\omega
 _3 (t-T_2)/2}
\end{cases}
\end{align*}
with 
\begin{align*}
t \in 
\begin{cases}
& [0,T_1], \\ 
& [T_1 , T_2], \\
& [T_2 , T],
\end{cases} \ 
\begin{cases} 
\omega _1 &= qB_1/m, \\ 
\omega _2 &= qB_2/m, \\
\omega _3 &= qB_3 /m, 
\end{cases} \ 
\begin{cases}
\alpha_1 = \zeta _1 (T_2), \\ 
\alpha _2 = \zeta _1'(T_2),
\end{cases} \ 
\begin{cases}
\beta_1 = \zeta _2 (T_2), \\ 
\beta _2 = \zeta _2'(T_2).
\end{cases}
\end{align*}
Then 
\begin{align*}
 \zeta _1(T) + \zeta _2'(T) = 
(\alpha _1 + \beta _2) \COS{\omega _3 (T-T_2)/2} + (2 \alpha _1 / \omega
 _3 - \omega _3 \beta _1 /2) \SIN{\omega _3 (T-T_2)/2}
\end{align*}
holds. By denoting 
\begin{align*}
& c_1 = \COS{\omega _1 T_1/2}, \quad c_2 = \COS{\omega _2 (T_2-T_1)/2},
 \quad 
c_3 = \COS{\omega _3 (T-T_2)/2}, \\
& s_1 = \SIN{\omega _1 T_1/2}, \quad s_2 = \SIN{\omega _2 (T_2-T_1)/2},
 \quad 
s_3 = \SIN{\omega _3 (T-T_2)/2}, 
\end{align*}
we have 
\begin{align*}
D =  \zeta _1(T) + \zeta _2'(T) = \left( 
2c_1c_2 - \frac{\omega _1^2 + \omega _2 ^2}{\omega _1 \omega _2} s_1s_2
\right) c_3 - \left( 
\frac{\omega _2^2 + \omega _3^2}{\omega _3 \omega _2} c_1 s_2 +
 \frac{\omega _1^2 + \omega _3^2}{\omega _1 \omega _3} s_1 c_2
\right)s_3.
\end{align*}
It is difficult to deduce a general condition such that
Assumption \ref{A1} holds. Thus, we see an example. Set 
$$
c_1=c_2=c_3=s_1=s_2=s_3 = 1/ \sqrt{2},
$$ 
and then 
$$ 
D \leq 2c_1c_2c_3 - 2s_1s_2c_3-2c_1s_2s_3-2s_1c_2s_3 = -2 \sqrt{2 } < -2
$$
holds if $\omega_1 \omega _2 >0$, $\omega _3 \omega _2 >0$ and $\omega_1
\omega _3 >0$ hold. Furthermore 
\begin{align*}
\zeta _2(T) &= (1/\sqrt{2}) (1/ \omega _1 + 1/ \omega _2 + 1 / \omega _3 - \omega
 _2/(\omega _1 \omega _3)) 
\end{align*}
holds. Then, for example, by putting $\omega _1 =  \omega _2 =  \omega _3 $, we have $\zeta _2(T) \neq 0$.





\end{document}